\documentclass[conference]{IEEEtran}
\IEEEoverridecommandlockouts

\usepackage[T1]{fontenc}
\usepackage[utf8]{inputenc}
\usepackage{lmodern}
\usepackage{microtype}

\usepackage{amsmath,amssymb,amsfonts}
\usepackage{mathtools}

\usepackage{booktabs}
\usepackage{array}
\usepackage{tabularx}
\usepackage{multirow}
\usepackage{makecell}

\usepackage{graphicx}
\usepackage{subcaption}

\usepackage{url}
\usepackage{hyperref}
\hypersetup{
  colorlinks=true,
  linkcolor=blue!50!black,
  citecolor=blue!50!black,
  urlcolor=blue!50!black
}
\usepackage{xcolor}
\usepackage{textcomp}
\usepackage{orcidlink}

\newcommand{\Nmax}{N_{\max}}
\newcommand{\dI}{\Delta I}
\newcommand{\mprismsub}{\ensuremath{\text{PRISM}_{\text{sub}}}}
\newcommand{\membersub}{\ensuremath{\text{EMBER}_{\text{sub}}}}
\newcommand{\mprismfull}{\ensuremath{\text{PRISM}_{\text{full}}}}
\newcommand{\mprismpool}{\ensuremath{\text{PRISM}_{\text{pool}}}}
\newcommand{\mprismtemp}{\ensuremath{\text{PRISM}_{\text{temporal}}}}

\begin{document}

\title{PRISM: PE Relational Inter-Section Matrix\\
A 2D Section-Aware Dataset for Static PE Malware Detection}

\author{
\IEEEauthorblockN{José M. Sacristán\IEEEauthorrefmark{1}\,\orcidlink{0009-0001-6621-8219}}
\IEEEauthorblockA{Universidad Carlos III de Madrid (UC3M)\\
\texttt{josemanuel.sacristan@alumnos.uc3m.es}}\and
\IEEEauthorblockN{Ana I. González-Tablas\,\orcidlink{0000-0002-6259-8955}}
\IEEEauthorblockA{Universidad Carlos III de Madrid (UC3M)\\
\texttt{aigonzal@inf.uc3m.es}}
}

\maketitle

\begin{abstract}
We introduce PRISM (PE Relational Inter-Section Matrix), an open
dataset and feature representation for static Windows PE malware
detection. Existing benchmarks such as EMBER, BODMAS, and
SOREL-20M represent each PE file as a flat one-dimensional feature
vector, discarding the ordering of sections and the relational
context between them. PRISM instead encodes every binary as a
two-dimensional matrix whose rows are individual PE sections in
file order, with a global summary row that preserves compatibility
with EMBER-style models. We build the corpus from four malware
sources (BODMAS, MalwareBazaar, VirusShare, and CAPE) together
with SOREL-20M benign software, yielding 83{,}633 deduplicated
matrices and a family-filtered analysis corpus of 49{,}204 samples
across 684 malware families.

A formal separability analysis (Fisher Discriminant Ratio, mutual
information, and inter-section information gain) shows that the
per-section positional structure carries discriminative
information that flat representations cannot capture. Under a
strictly controlled, sample-matched comparison, a gradient-boosted
classifier on the compact PRISM representation recovers nearly all
of the binary-detection performance of the same classifier on the
much larger EMBER vector, at roughly one-sixth the dimensionality;
EMBER retains only a small, consistent advantage confined to the
extreme low-false-positive regime, the two being operationally
indistinguishable at the decision threshold. We are explicit that
this binary task is saturated, so the structural content PRISM
preserves is reserved for tasks with greater metric headroom, such
as family classification and architectures that exploit the 2D
structure directly. The dataset, extraction library, trained
models, and full analysis pipeline are released under CC~BY~NC-SA
and MIT licences.
\end{abstract}


\section{Introduction}
\label{sec:intro}

The detection of malicious Windows Portable Executable (PE) files through static machine learning has become one of the most active and practically consequential problems in applied cybersecurity. Every day, commercial antivirus engines and endpoint detection platforms must classify millions of PE files --- executables, DLLs, and system drivers --- against an adversarial backdrop of constantly evolving malware families. The economic and operational stakes are substantial: a false negative in a corporate environment may allow a ransomware payload to encrypt thousands of files before behavioural detection triggers; a false positive in a high-throughput environment may block legitimate software updates and erode user trust.

Since the release of the EMBER dataset in 2018, the standard representation used by virtually every public benchmark has been a flat 1D feature vector of 2{,}381 dimensions, obtained by concatenating eight heterogeneous feature groups --- including byte histograms, string features, header fields, section statistics, and import/export information --- extracted using the LIEF library~\cite{anderson2018ember}. This representation was a significant contribution to the field: it standardised feature extraction, enabled reproducible comparisons across research groups, and provided a large enough corpus to train competitive gradient-boosted models. The EMBER baseline LightGBM model achieved AUC-ROC 0.99338 and TPR@FPR=0.1\% of 0.8027 on the EMBER2018 benchmark. EMBER2018 is the harder of the two releases, deliberately constructed so that its train/test split is more difficult for machine-learning classifiers than the earlier EMBER2017 set; these figures have accordingly been accepted by much subsequent work as a representative operating point for static 1D approaches.

However, the EMBER representation discards two structural pieces of information that are present in every PE binary and are well understood by malware analysts: (1) the ordering of sections within the file, and (2) the relational context between consecutive or co-occurring sections. A packed executable, for instance, typically presents a high-entropy stub section at position 0, followed by one or more compressed or encrypted payload sections whose size ratios and permission flags differ systematically from those of legitimate binaries. When this structural pattern is aggregated into a single global vector, the positional signal is irreversibly lost. A classifier operating on the EMBER vector cannot distinguish a packer stub at position 0 from the same byte statistics occurring at position 5.

This observation is not merely theoretical. Our empirical analysis shows that the MEM\_DISCARDABLE memory permission flag at section position 5 (MEM\_DISC@SEC5, using the feature--position notation defined in Section~\ref{sec:prism-representation}) achieves a Fisher Discriminant Ratio (FDR) of 1.287 --- where FDR measures inter-class separability as the ratio of between-class variance to within-class variance, formally defined in Section~\ref{sec:fdr} --- the single most discriminative (section, feature) cell in our entire feature space. The maximum FDR attained by any single dimension of the PRISM global summary row is 0.858 (\texttt{log\_exports}), so the per-section positional representation attains a $1.50\times$ higher peak FDR than any file-level descriptor on the same corpus. More broadly, 12{,}854 inter-section feature pairs across the (section $\times$ feature) lattice carry non-trivial additional mutual information about the malware label ($\dI > 0.01$ bits) beyond either component feature alone, with the top pair --- (raw\_size@SEC2, name5@SEC3) --- contributing $\dI = 0.205$ bits of additional discriminative information beyond any individual cell. These findings motivate a fundamentally different representation: one that preserves section identity, ordering, and relational context as first-class dimensions.

A second motivation for the present release is temporal. BODMAS~\cite{yang2021bodmas}, the most recent publicly available temporal benchmark, covers malware collected between August 2019 and September 2020. In the five years since that collection window closed, the malware landscape has changed substantially: new families such as LummaStealer, AsyncRAT, DarkGate, and Rhadamanthys have emerged and proliferated; the prevalence of commodity stealers and loaders distributed through malware-as-a-service platforms has increased; and evasion techniques targeting ML-based classifiers have become more sophisticated. PRISM is the first open benchmark to include PE malware samples from 2024--2025 collected from MalwareBazaar alongside re-processed BODMAS data, enabling evaluation against the contemporary threat environment.

The contributions of this paper are:

\begin{itemize}
\item \textbf{PRISM representation.} We define a 2D matrix $M \in \mathbb{R}^{(\Nmax+1)\times F}$ that encodes each PE section as a row with $F = 25$ semantic features covering name encoding, size ratios, permission flags, entropy and entropy quartiles, positional index, and structural anomaly flags. An additional global summary row provides backward compatibility with EMBER-style models.

\item \textbf{Multi-source contemporary corpus.} We process malware binaries from BODMAS~\cite{yang2021bodmas} (2019--2020), MalwareBazaar (2024--2025), VirusShare collection 00499, and historical CAPE sandbox samples, together with 29{,}467 benign executables from SOREL-20M, yielding 83{,}633 unique PRISM matrices after deduplication. A family-filtered primary corpus of 49{,}204 samples (19{,}737 malware with verified family labels across 684 families, plus 29{,}467 benign) supports all separability and baseline analyses.

\item \textbf{Formal separability analysis.} We provide a systematic measurement of Fisher Discriminant Ratio (FDR), Mutual Information (MI), and inter-cell information gain ($\dI$) at the (section index, feature type) granularity on the full 49{,}204-sample family-filtered corpus, demonstrating that the most discriminative positional feature (MEM\_DISC@SEC5, FDR = 1.287) achieves a $1.50\times$ higher FDR than any single dimension of the PRISM global summary row, and that 12{,}854 inter-section feature pairs in the (section $\times$ feature) lattice carry non-trivial information gain ($\dI > 0.01$ bits) beyond any individual feature.

\item \textbf{Five reference model configurations including a controlled cross-representation comparison.} We train LightGBM~\cite{ke2017lightgbm} models under a strictly controlled protocol with explicit deduplication and zero train/test overlap: $\mprismpool$ (PRISM per-section features mean-pooled across sections, position discarded, 25-dim), $\mprismfull$/$\mprismsub$ (PRISM flattened, 425-dim, on the full corpus and on the EMBER-compatible sub-corpus respectively), $\mprismtemp$ (PRISM flattened under a BODMAS-malware temporal split, 425-dim), and $\membersub$ (EMBER 1D, 2{,}381-dim, compared against $\mprismsub$ on identical samples and splits). Including BODMAS via restored disarmed binaries, the cross-representation comparison over 20 deterministic splits shows that PRISM (425-dim) attains binary-detection performance close to EMBER (2{,}381-dim) at one-sixth the dimensionality, conceding only a small, consistent gap ($+0.85$ pp TPR@FPR=0.1\% in EMBER's favour) confined to the deep-FPR tail, while remaining operationally indistinguishable at the decision threshold.

\item \textbf{Open release.} The dataset, the \texttt{prism-extract} extraction library, all baseline models, the controlled benchmark code, and the complete analysis pipeline are released under CC BY 4.0 and MIT licences respectively.
\end{itemize}

The remainder of this paper is organised as follows. Section~\ref{sec:related} surveys related datasets and feature representations. Section~\ref{sec:prism-representation} formally defines the PRISM matrix and per-section feature set. Section~\ref{sec:dataset} describes the dataset construction pipeline. Section~\ref{sec:separability} presents the separability analysis. Section~\ref{sec:baselines} defines the baseline experimental configurations. Section~\ref{sec:results} reports and interprets the results. Section~\ref{sec:comparison} compares PRISM against existing datasets. Sections~\ref{sec:conclusion} and~\ref{sec:future} outline conclusions and future directions.

\section{Related Work}
\label{sec:related}

\subsection{Public PE Malware Datasets}

The availability of large, labelled, and reproducible PE malware datasets has been a critical enabler of progress in static malware detection research. Prior to 2018, most published work relied on private or proprietary datasets, making comparison across papers unreliable and entry barriers for new researchers high.

Anderson and Roth~\cite{anderson2018ember} addressed this gap with EMBER, the first large-scale open benchmark for static PE malware detection. EMBER provides pre-extracted features for 1.1 million PE files scanned in or before 2018, together with an open-source LIEF-based feature extractor and a LightGBM baseline model. The 2{,}381-dimensional EMBER feature vector concatenates byte histogram, string features, general file information, header features, section information, and data directory fields into a single flat representation. The EMBER vector has since become the de facto standard for PE malware feature engineering and the original paper one of the most heavily cited resources in the field. Joyce et al.~\cite{joyce2025ember2024} recently released EMBER2024, extending the corpus to 3.2 million files across six file formats and introducing a ``challenge set'' of malware that initially evaded antivirus detection. However, EMBER2024 retains the flat 1D vector philosophy and does not expose section-level positional structure.

Yang et al.~\cite{yang2021bodmas} introduced BODMAS, adding two critical dimensions absent from EMBER: temporal labels and malware family annotations. BODMAS covers 57{,}293 malware samples and 77{,}142 benign samples collected between August 2019 and September 2020, organised chronologically so that researchers can study temporal concept drift. The BODMAS paper demonstrated that models trained on earlier samples lose significant detection capability when evaluated on later samples, motivating the temporal split evaluation we adopt for our $\mprismtemp$ model (Section~\ref{sec:b3}). BODMAS malware is distributed in \emph{disarmed} form --- with the PE \texttt{Machine} and \texttt{Subsystem} header fields zeroed to prevent accidental execution --- rather than as feature vectors only; we show in Section~\ref{sec:b4} that these binaries can be restored and processed by the EMBER extractor, enabling BODMAS to be included in the controlled cross-representation benchmark. Abdulwahab et~al.~\cite{abdulwahab2026realtime} confirmed 
this ranking on EMBER~2024, finding that LightGBM and XGBoost 
(AUC-ROC\,=\,0.9979) substantially outperform MLP and 
TabNet-based architectures on the same 2{,}381-dimensional 
tabular feature set, with all pairwise differences 
statistically significant by McNemar's test.

Harang and Rudd~\cite{harang2020sorel} introduced SOREL-20M (Sophos/ReversingLabs-20 Million), scaling the public PE benchmark to near-industrial size with approximately 20 million samples, pre-extracted features, and approximately 10 million ``disarmed'' binaries --- samples with executable headers zeroed to prevent accidental execution, following a methodology we adopt for the BODMAS portion of our corpus. SOREL-20M addressed the training-size limitations of EMBER but is computationally demanding and, like EMBER, provides only 1D feature vectors. The benign subset of SOREL-20M is the source of the benign corpus used in the present release. Yousuf et al.~\cite{yousuf2022multifeature} extended the feature set by constructing a dataset that includes not only PE header and section features but also DLL names and imported function lists. Wang et al.~\cite{wang2022measurement} provided a large-scale empirical study of malware family classification, finding that family-level labels from antivirus tools are substantially noisy and that this noise significantly affects benchmark validity --- a finding relevant to our family-filtered corpus protocol.

None of the datasets described above preserve section ordering as a first-class structural dimension. All represent each PE file as a single fixed-length vector, discarding the sequential and relational structure of the section table that is central to the PRISM design.

\subsection{Static Feature Representations}

Feature extraction strategies for static PE analysis span a broad spectrum from raw bytes to semantically engineered feature vectors. At one extreme, Rezaei et al.~\cite{rezaei2021peheader} demonstrate that as few as 324 bytes of raw PE header content --- comprising the DOS header, File Header, and Optional Header --- contain sufficient discriminative information to train a deep embedding model achieving over 97\% accuracy on balanced datasets. Their use of a joint deep embedding and k-means clustering objective is particularly relevant to PRISM's design philosophy: they show that unsupervised structural groupings in the embedding space align with malware/benign labels, suggesting that the PE header encodes latent structural regularities that go beyond simple feature engineering.

Nakro\v{s}is et al.~\cite{nakrosis2022pe} apply convolutional networks directly to PE header byte sequences, treating the header as a one-dimensional signal and learning positional filters that capture local byte patterns. Their approach achieves competitive accuracy without manual feature design, but at the cost of interpretability: the learned filters do not correspond to named fields or section boundaries. Maleki et al.~\cite{maleki2019packed} focus specifically on packed malware detection through analysis of the PE section table --- the closest precursor to the PRISM representation. They observe that packers introduce characteristic anomalies in section sizes, virtual-to-raw size ratios, and permission flag combinations, and show that these anomalies can be detected by analysing the section table structure rather than byte content. PRISM generalises this insight by encoding section-table features systematically across all sections and all feature dimensions, enabling both traditional ML and deep learning approaches to exploit the full relational structure of the section table.

Yuk and Seo~\cite{yuk2022pe} and Yousuf et al.~\cite{yousuf2023peerj} provide complementary studies of combinations of PE header fields and section statistics using traditional ML classifiers, confirming that section-level features consistently outperform header-only features in binary classification tasks. Hasanah et al.~\cite{hasanah2025review} provide a recent systematic review of ML approaches to malware detection, identifying the lack of temporally diverse, openly available datasets as a key bottleneck for progress. All of the approaches surveyed, however, aggregate section-level information into global statistics before classification, discarding the positional context that PRISM preserves.

\subsection{Image-Based and Hybrid Representations}

An orthogonal line of research converts PE binary content into greyscale images and applies CNN-based vision classifiers. Malik et al.~\cite{malik_matin_image_2025} demonstrate that transfer learning from ImageNet-pretrained CNNs to malware images achieves competitive classification accuracy on standard benchmarks. Hai et al.~\cite{hai_proposed_2023} integrate an image-based malware detector (using MobileNetV2 and InceptionV3 fine-tuned on BODMAS) into a full endpoint detection and response system, showing that lightweight CNNs can achieve AUC above 0.86 at inference times below 0.1 seconds per file. Yu et al.~\cite{yu_semantic_2025} and Zhao et al.~\cite{zhao_mcpds_2026} explore more structured image representations that partially preserve section boundaries.

While image-based approaches implicitly preserve some spatial structure, they do so without explicit section boundary semantics: a pixel at position (100, 50) in a malware image does not correspond to a named section or a specific feature dimension. This makes it difficult to apply interpretability tools such as Fisher Discriminant Ratio or Mutual Information at the section-feature granularity, and prevents direct compatibility with EMBER-style tabular models. El-Hajj~\cite{el-hajj_hybrid_2025} combines static and dynamic features in a hybrid approach, noting that pure static analysis is insufficient for highly obfuscated malware and that dynamic features provide complementary signal. PRISM is designed as a static analysis representation, but its 2D structure is architecturally compatible with the addition of a third dimension for dynamic features --- a direction we pursue in planned future work.

\subsection{Separability and Information Theory}

Kraskov et al.~\cite{kraskov_estimating_2004} introduced the k-nearest-neighbour estimator for Mutual Information that we use in Section~\ref{sec:separability}, providing a non-parametric estimator that avoids the binning artefacts of histogram-based methods and is well-suited to high-dimensional continuous features. The consistent superiority of gradient-boosted classifiers 
over simpler models on EMBER-scale tabular PE 
features~\cite{anderson2018ember,yang2021bodmas,abdulwahab2026realtime} 
motivates our use of LightGBM as the classifier family for 
all five baseline configurations. Our contribution extends this line of analysis to the (section index, feature type) granularity, asking not merely which classifier family is most effective but which pairs of (section, feature) cells across the matrix carry information beyond their individual components. This is, to our knowledge, the first systematic application of inter-cell Mutual Information gain analysis to PE malware datasets.

\section{The PRISM Representation}
\label{sec:prism-representation}

\subsection{Formal Definition}

Let a PE binary contain $N$ sections ($N \geq 1$). We define its PRISM matrix as
\begin{equation}
M \in \mathbb{R}^{(\Nmax+1)\times F}
\end{equation}
where $\Nmax = 16$ is the maximum number of sections considered (the empirical 95th-percentile of section counts across the corpus is $P_{95} = 9$, meaning that $\Nmax = 16$ covers 100\% of all samples in the family-filtered corpus while providing margin for outlier binaries --- the full justification and section-count distribution are presented in Section~\ref{sec:design-rationale}), and $F = 25$ is the effective number of features per row used in the analysis.

Rows $0, \ldots, N-1$ correspond to the $N$ real PE sections in file order. Row $\Nmax$ is a global summary row compatible with EMBER-style models. Rows $N, \ldots, \Nmax-1$ are zero-padded, with a binary mask vector $m \in \{0,1\}^{\Nmax+1}$ indicating active rows. This design choice --- a fixed-size matrix with explicit masking --- allows PRISM matrices to be stacked into tensors for batch processing by deep learning frameworks without dynamic padding.
\subsection{Per-Section Feature Vector}
\label{sec:per-section-features}

Each section row $M_i$, $i < \Nmax$, is composed of seven feature groups (Table~\ref{tab:features}).

\begin{table}[t]
\centering
\caption{PRISM per-section feature set. Total $F = 25$ dimensions per section row.}
\label{tab:features}
\begin{tabularx}{\columnwidth}{@{}l X c@{}}
\toprule
\textbf{Group} & \textbf{Features} & \textbf{Dim.} \\
\midrule
Name enc.   & Section name hashed to 8-dim bin vector & 8 \\
\addlinespace
Sizes       & SizeOfRawData, VirtualSize, raw/virt ratio (log-scale) & 3 \\
\addlinespace
Permissions & READ, WRITE, EXEC, DISC, CODE, DATA flags & 6 \\
\addlinespace
Entropy     & Shannon entropy $H$ of section byte content & 1 \\
\addlinespace
Quartiles   & Q2--Q4 entropy over 256-byte sliding windows & 3 \\
\addlinespace
Position    & Normalised section index $i / \Nmax$ & 1 \\
\addlinespace
Anomaly     & Unusual name; WX co-occurrence; zero raw size & 3 \\
\midrule
\textbf{Total} & & \textbf{25} \\
\bottomrule
\end{tabularx}
\end{table}

\textit{Note on feature schema.} An earlier internal version of the PRISM extractor included a first quartile (Q1) of windowed entropy as a 26th feature; following a structural redundancy analysis, this slot was removed because Q1 exhibits empirical correlation above 0.95 with both the byte entropy and Q2 features across the corpus, contributing no additional discriminative information. This compresses the per-section feature allocation from $F = 26$ to $F = 25$, yielding the $425$-dimensional flattened representation 
($(N_{\max}+1) \times F = 17 \times 25$) used in all baseline experiments. The reference implementation in \texttt{prism-extract} retains a reserved slot at index 22 for backward compatibility with previously-extracted matrices, populated with zeros. All analyses in this paper operate on these $F = 25$ effective feature dimensions; this is reflected in the LightGBM input dimensionality reported in Section~\ref{sec:baselines}.

\subsection{Global Row}

Row $\Nmax$ encodes file-level summary features: normalised section count, log-scaled import/export counts, digital signature presence, and resource section presence. This row provides a self-contained EMBER-compatible global descriptor; it is included as part of the flattened PRISM vector (Section~\ref{sec:b2}) and serves as one comparison arm --- the per-section maximum versus the global-row maximum --- in the separability analysis of Section~\ref{sec:separability}. Of the 25 feature slots in the global row, 5 are populated with these file-level descriptors and 20 are zero by construction.

\subsection{Design Rationale}
\label{sec:design-rationale}

\begin{figure}[t]
\centering
\includegraphics[width=\columnwidth]{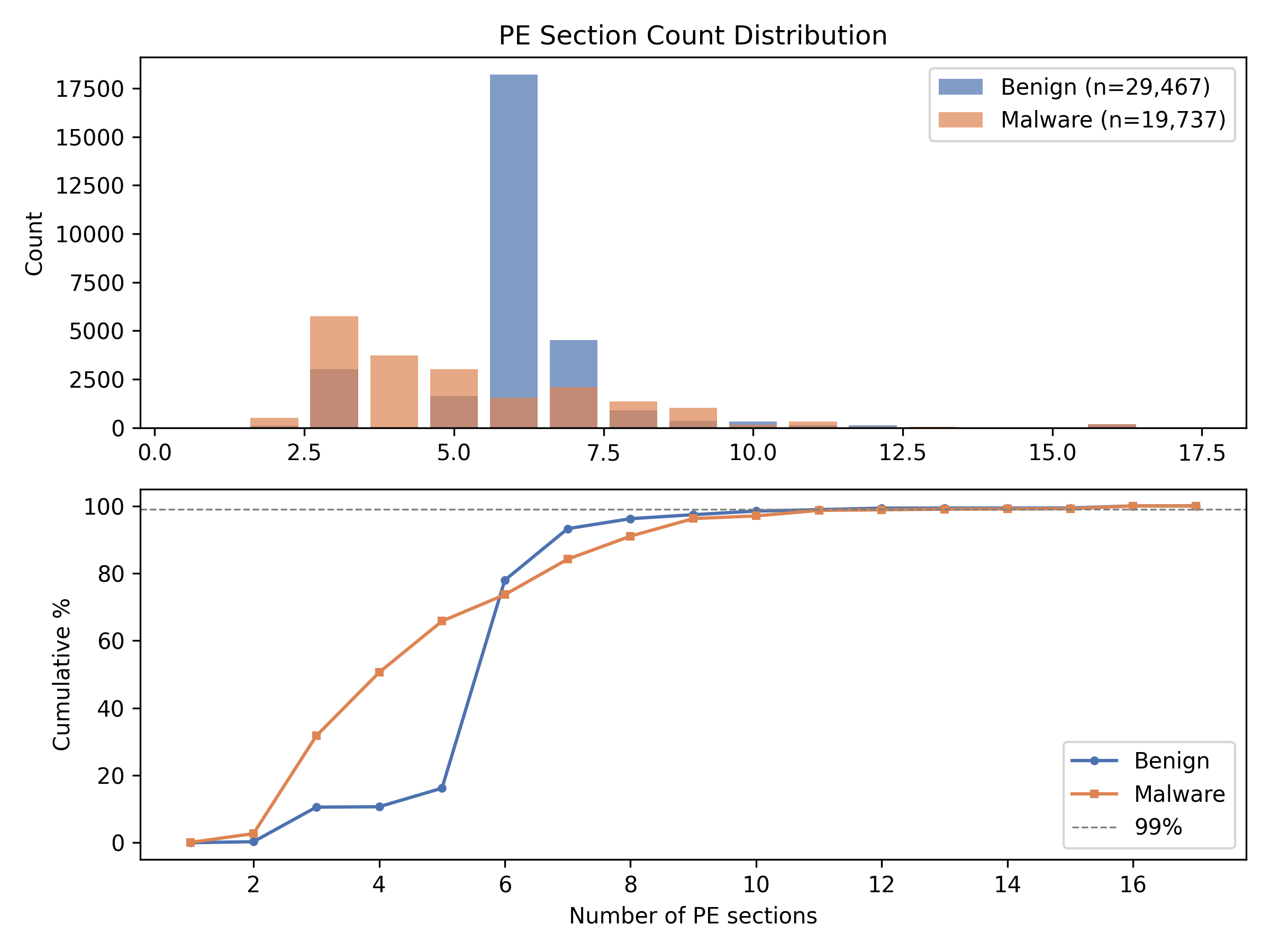}
\caption{Distribution of PE section counts by class on the 49{,}204-sample family-filtered corpus (29{,}467 SOREL benign, 19{,}737 family-labelled malware). Malware (red) shows a bimodal distribution with a primary peak at 3 sections and a secondary peak at 7 (section count 6 is a local minimum). Benign (blue) is sharply concentrated at 6 sections. Dashed lines indicate corpus-wide percentiles: P50 = 6, P90 = 8, P95 = 9, P99 = 12. Setting $\Nmax = 16$ covers 100\% of the corpus.}
\label{fig:section-distribution}
\end{figure}

The choice of $\Nmax = 16$ is justified empirically by the section count distribution observed in our corpus. As shown in Figure~\ref{fig:section-distribution}, the 50th percentile of section counts is 6, the 90th percentile is 8, the 95th percentile is 9, and the 99th percentile is 12. Setting $\Nmax = 16$ therefore covers 100\% of all samples in the family-filtered corpus without truncation, while keeping the matrix dimensions compact enough for efficient batch processing. The malware distribution is bimodal: a large proportion of samples have only 2--3 sections (consistent with simple packers that compress the original binary into a single payload section), while a tail extends to higher section counts (consistent with modular malware or installers with multiple resource sections). The benign distribution, drawn from SOREL-20M, is much more concentrated, with the majority of samples having exactly 6 sections.

The feature set was designed to be extractable using LIEF~\cite{thomas_lief_2017} from any PE binary in under 50 ms on commodity hardware, enabling large-scale corpus construction without GPU resources. All 25 features are either directly available as LIEF attributes or computable from raw section byte content (entropy, quartiles), making the pipeline fully reproducible from the released \texttt{prism-extract} library.

\section{Dataset Construction}
\label{sec:dataset}

\subsection{Malware Sources}
\label{sec:malware-sources}

The PRISM corpus draws from four complementary malware repositories, chosen to maximise temporal and family diversity while remaining fully reproducible from publicly accessible sources.

\begin{figure}[t]
\centering
\includegraphics[width=\columnwidth]{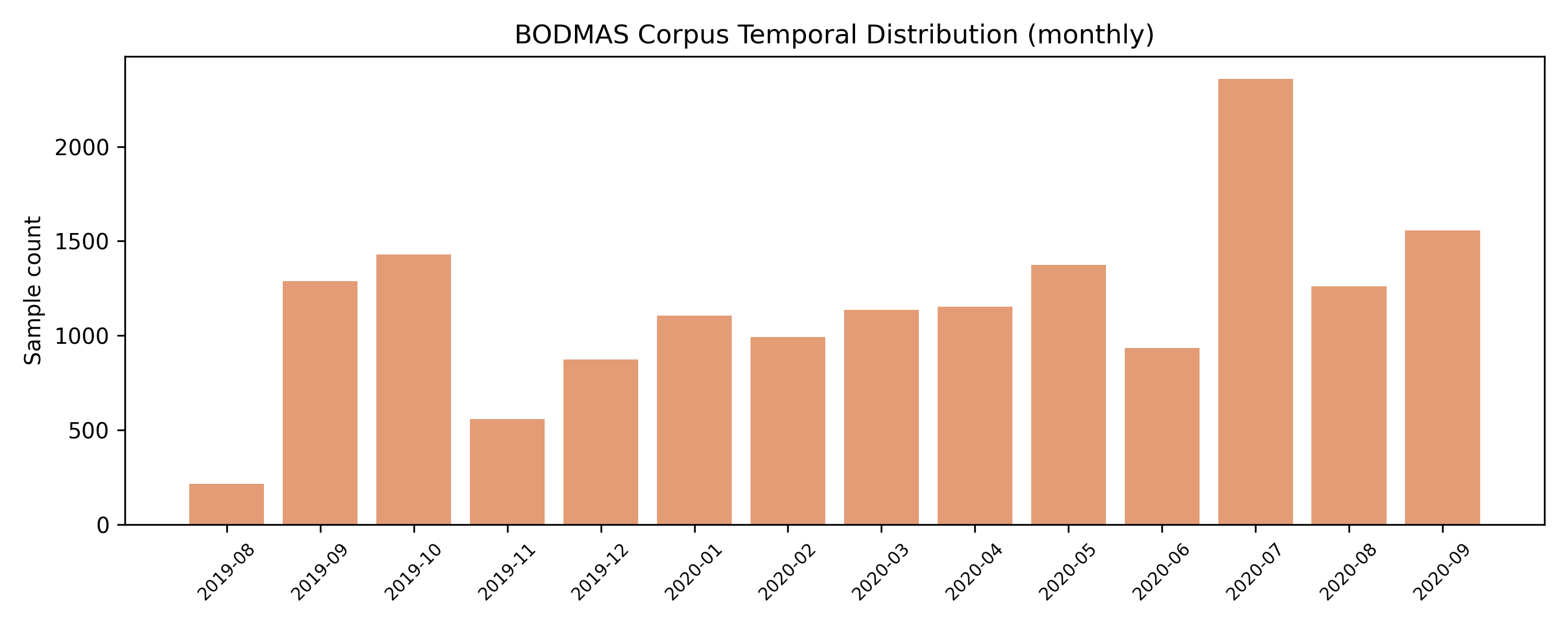}
\caption{Monthly distribution of BODMAS samples between August 2019 and September 2020. The split point used for the $\mprismtemp$ single-class temporal probe (Section~\ref{sec:b3}) is the 80th-percentile first-seen timestamp of the BODMAS malware, 2020-07-25.}
\label{fig:bodmas-temporal}
\end{figure}

\begin{figure}[t]
\centering
\includegraphics[width=\columnwidth]{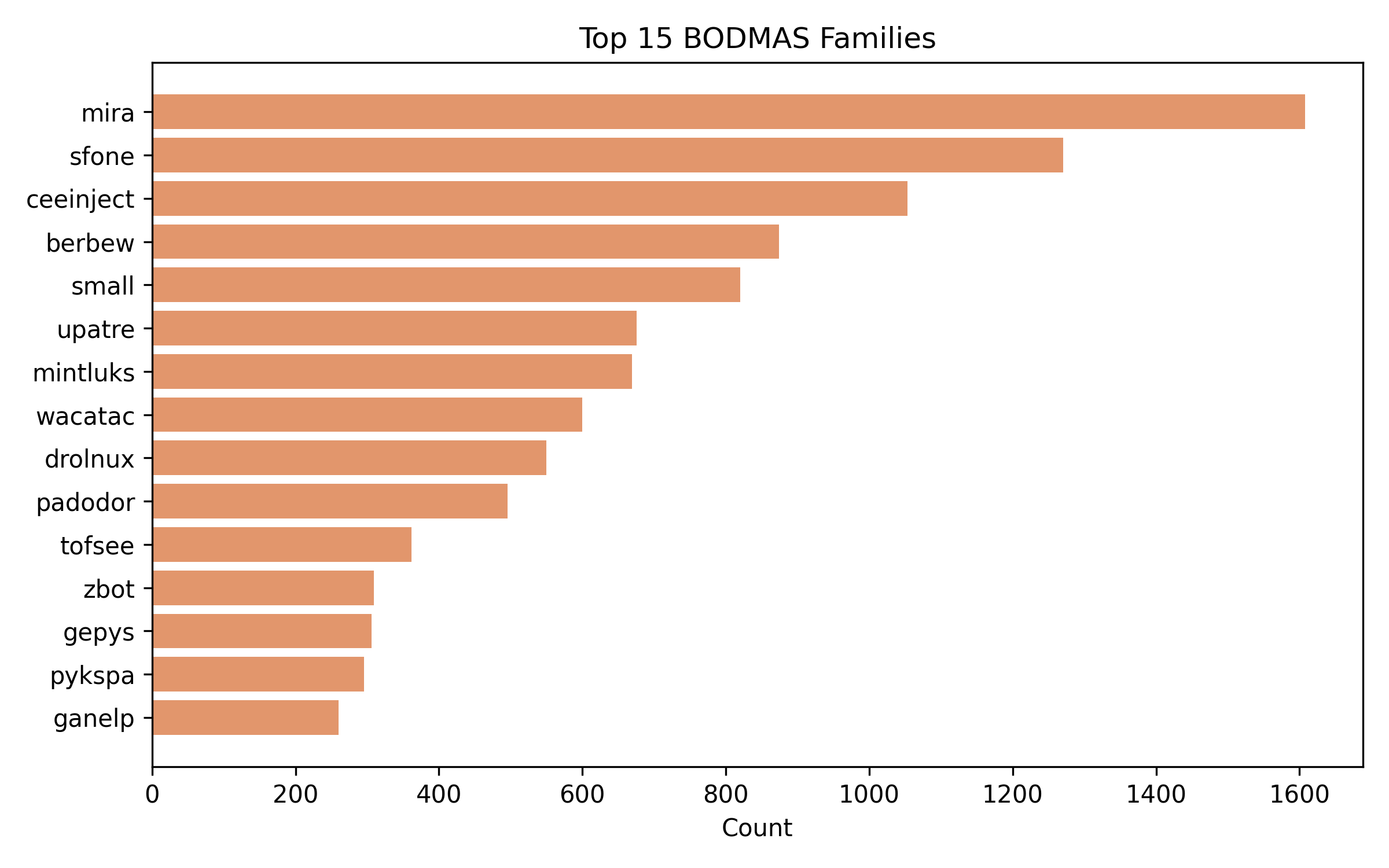}
\caption{Top 15 malware families in BODMAS by sample count. The corpus is dominated by commodity families (mira, sfone, ceeinject, berbew, small), which collectively account for over 30\% of all BODMAS malware samples.}
\label{fig:bodmas-families}
\end{figure}

\textbf{BODMAS}~\cite{yang2021bodmas} contributes disarmed PE malware samples collected between August 2019 and September 2020. Of the 57{,}218 disarmed binaries we obtained, 57{,}133 were successfully processed into PRISM matrices (85 failed extraction); these span 538 malware families after family filtering. The temporal distribution of BODMAS is shown in Figure~\ref{fig:bodmas-temporal} and the family distribution in Figure~\ref{fig:bodmas-families}. Binaries are distributed in disarmed form with the PE \texttt{Machine} and \texttt{Subsystem} header fields zeroed, following the SOREL-20M~\cite{harang2020sorel} methodology; this does not affect the PE section structure on which PRISM relies and is fully compatible with LIEF parsing. Unlike the feature-vector-only releases, the availability of these binaries lets us restore the zeroed header fields (Section~\ref{sec:b4}) and extract native EMBER vectors, so that BODMAS can be included in the controlled cross-representation benchmark (specifically, in its BODMAS-inclusive confirmatory run; see Section~\ref{sec:b4}).

\textbf{MalwareBazaar}~\cite{abusech_malwarebazaar_2020} contributes contemporary 2024--2025 PE samples obtained via the abuse.ch public API. Samples were retrieved by querying multiple malware family signatures (including LummaStealer, RemcosRAT, AsyncRAT, Emotet, DCRat, and NjRAT) and behavioural tags (ransomware, stealer, loader, RAT, backdoor, banker, dropper, botnet). This source is the contemporary anchor of the PRISM corpus, addressing the temporal gap identified in Section~\ref{sec:intro}.

\textbf{VirusShare}~\cite{godwin_virusshare_2012} collection 00499 contributes PE binaries identified by MZ magic byte detection from a broader file collection that included Android APKs, PDF documents, Office files, and other non-PE formats. VirusShare samples are included in the full deduplicated corpus but \emph{excluded from the family-filtered primary analysis corpus}, because VirusShare collection 00499 does not provide per-sample family labels.

\textbf{CAPE historical samples} were extracted from PRISM matrices computed during previous internal CAPE sandbox runs, restricted to the static slice of each tensor. Family labels were available for the subset of CAPE samples for which prior VirusTotal annotation existed; samples without family labels were excluded from the family-filtered corpus on the same basis as VirusShare.

\subsection{Benign Source}

Benign PE files are drawn from the SOREL-20M benign distribution~\cite{harang2020sorel}, yielding 29{,}467 unique benign matrices after deduplication. SOREL benigns are distributed in disarmed form analogous to the SOREL malware portion, preserving the PE section table for LIEF parsing. This source provides a heterogeneous benign distribution spanning multiple compilers, toolchains, and source archives, supporting separability and classification analyses with broad benign coverage.

\subsection{Extraction Pipeline}

We implemented the PRISM extractor (\texttt{prism-extract}) using LIEF 0.14.1~\cite{thomas_lief_2017} on Ubuntu 24.04 with Python 3.12. The extraction pipeline processes each PE file in under 50 ms on average on an 8-core VM with NVMe storage. The full deduplicated corpus and the family-filtered corpus statistics are reported in Section~\ref{sec:dedup}.

\subsection{Deduplication and Quality Control}
\label{sec:dedup}

The combined raw collection comprised 178{,}740 candidate PRISM matrices across all sources. Exact deduplication by feature matrix hash, using \texttt{np.unique} on flattened rows (ensuring reproducibility across runs), removed 53.2\% of the matrices, reflecting the substantial overlap between repositories that draw from shared threat intelligence feeds; BODMAS, MalwareBazaar, and VirusShare all index widely distributed malware families within hours of first appearance, resulting in extensive cross-source duplication. After deduplication, 83{,}633 unique matrices remained. A family filter was then applied to the malware portion: only samples carrying a verified family label from their source metadata (BODMAS family CSV, MalwareBazaar query tag, or per-sample VirusTotal annotation for CAPE) were retained for the primary analysis corpus. The final composition is summarised in Table~\ref{tab:corpus}. We note that exact-matrix-hash deduplication removes only byte-identical PRISM representations; it does not remove near-duplicate polymorphic variants whose matrices differ in a single cell. Residual near-duplicate leakage of this kind would inflate the absolute detection metrics of the random-split models ($\mprismpool$, $\mprismfull$); the $\mprismtemp$ temporal probe (Section~\ref{sec:b3}) is included precisely to test whether such within-window leakage drives the observed saturation, and finds that it does not.

\begin{table}[t]
\centering
\caption{PRISM corpus composition after deduplication and family 
filtering. Each matrix has shape $17 \times 25$. The 
family-filtered corpus (49{,}204 samples) is used for all 
separability and classification analyses in 
Sections~\ref{sec:separability}--\ref{sec:results}.}
\label{tab:corpus}
\begin{tabular}{@{}lr@{}}
\toprule
\textbf{Stage} & \textbf{Count} \\
\midrule
Raw matrices combined (before deduplication) & 178{,}740 \\
Unique matrices after global deduplication   & 83{,}633  \\
\addlinespace
\textbf{Family-filtered primary analysis corpus} & \textbf{49{,}204} \\
\hspace{1em}Family-labelled malware (49{,}204 subset) & 19{,}737 \\
\hspace{1em}SOREL benign (49{,}204 subset)            & 29{,}467 \\
\addlinespace
Distinct malware families            & 684    \\
Malware:benign ratio (family-filtered) & 1 : 1.49 \\
\bottomrule
\end{tabular}
\end{table}

The full deduplicated corpus ($n = 83{,}633$) is released alongside the family-filtered corpus for researchers wishing to train classifiers that do not require family-level annotations; all primary analyses in Sections~\ref{sec:separability}--\ref{sec:results} operate on the 49{,}204-sample family-filtered subset.

\subsection{Evaluation Subsets}
\label{sec:subsets}

All experiments in this paper draw from the single 49{,}204-sample family-filtered corpus defined above. Three subsets of it are used, distinguished only by which samples participate; the per-sample features are identical across all three. We name them here so that the experimental section (Section~\ref{sec:baselines}) can refer to datasets and representations separately from the models trained on them.

\begin{enumerate}
\item \textbf{Full family-filtered set} ($n = 49{,}204$): the entire corpus (19{,}737 malware from BODMAS, CAPE, and MalwareBazaar; 29{,}467 SOREL benign). Used whenever no EMBER vector or timestamp constraint applies.

\item \textbf{Temporal-malware subset}: the same corpus, but with the BODMAS malware ordered by first-seen timestamp so that a temporal train/test cut can be applied to \emph{that class only}. The benign class (SOREL) carries no usable timestamp and is therefore always taken in full and split at random. Only the BODMAS malware is partitioned in time; no benign sample is ever removed.

\item \textbf{EMBER-compatible subset} ($n = 32{,}623$): the subset for which native EMBER 2{,}381-dim vectors can be extracted from raw binaries --- SOREL benign (29{,}137), CAPE (3{,}110), and MalwareBazaar (376). BODMAS is absent from this subset in the primary form because its binaries require restoration first; once restored (Section~\ref{sec:b4}), BODMAS adds 16{,}202 EMBER-valid samples, giving a BODMAS-inclusive variant of $48{,}825$ samples used as a confirmatory check. This subset is the only one on which the PRISM and EMBER representations can be compared on identical samples.
\end{enumerate}

We emphasise that these are \emph{subsets and representations}, not models. A model (always the same LightGBM, Section~\ref{sec:baselines}) becomes a named model only once it is trained on one of these subsets under one representation; the full-corpus and EMBER-subset PRISM models, although they share the flattened PRISM representation, are distinct models trained on distinct sample sets and are reported as such.

\begin{figure}[t]
\centering
\includegraphics[width=\columnwidth]{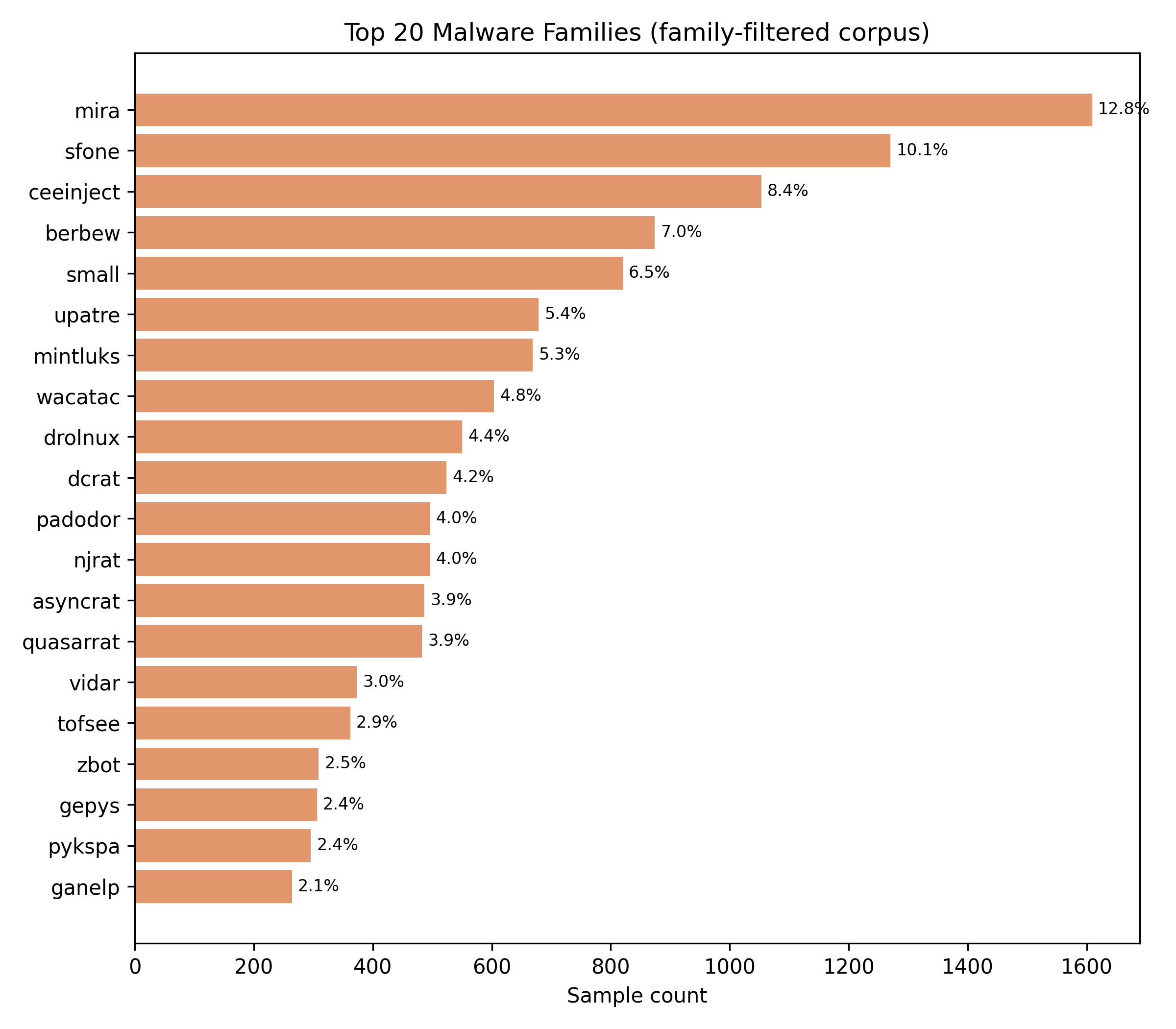}
\caption{Top 20 malware families in the family-filtered corpus by sample count. The corpus is dominated by mira (12.8\%), sfone (10.1\%), ceeinject (8.4\%), berbew (7.0\%), and small (6.5\%), which collectively account for 44.8\% of the family-labelled malware. The top 30 families cover 73.6\% of family-labelled samples; the remaining 26.4\% is distributed across 654 long-tail families.}
\label{fig:top-families}
\end{figure}

\section{Separability Analysis}
\label{sec:separability}

We perform three complementary analyses to quantify the discriminative content of the per-section positional representation. All analyses operate on the full 49{,}204-sample family-filtered corpus.

\subsection{Fisher Discriminant Ratio}
\label{sec:fdr}

The Fisher Discriminant Ratio (FDR) measures the separability between two classes for a given feature as the ratio of between-class variance to within-class variance. We apply FDR not to individual features in isolation but to each (section index $i$, feature $f$) pair independently:
\begin{equation}
\mathrm{FDR}(i, f) = \frac{(\mu_1(i, f) - \mu_0(i, f))^2}{\sigma^2_1(i, f) + \sigma^2_0(i, f)}
\end{equation}
where $\mu_c$ and $\sigma^2_c$ denote the class-conditional mean and variance of feature $f$ at section position $i$ for class $c \in \{0\text{ (benign)}, 1\text{ (malware)}\}$. A high FDR value indicates that the feature takes systematically different values for malware and benign samples at that specific section position.

\begin{table}[t]
\centering
\caption{Separability metrics on the 49{,}204-sample family-filtered corpus, comparing per-section positional features against the PRISM global summary row. The global column reports values for the 5 file-level descriptors populated in the global row; the remaining 20 slots are zero by construction.}
\label{tab:separability}
\begin{tabularx}{\columnwidth}{@{}l >{\centering\arraybackslash}X >{\centering\arraybackslash}X c@{}}
\toprule
\textbf{Metric} & \makecell{\textbf{Per-section}\\\textbf{maximum}} & \makecell{\textbf{Global row}\\\textbf{maximum}} & \textbf{Ratio} \\
\midrule
Max FDR & \textbf{1.287} & 0.858 & \textbf{1.50$\times$} \\
        & {\footnotesize(\texttt{MEM\_DISC@SEC5})} & {\footnotesize(\texttt{log\_exports})} & \\
\addlinespace
Max MI (bits) & \textbf{0.522} & 0.357 & \textbf{1.46$\times$} \\
              & {\footnotesize(\texttt{virt\_size@SEC1})} & {\footnotesize(\texttt{log\_imports})} & \\
\bottomrule
\end{tabularx}
\end{table}

\begin{figure}[t]
\centering
\begin{subfigure}{\columnwidth}
  \centering
  \includegraphics[width=\columnwidth]{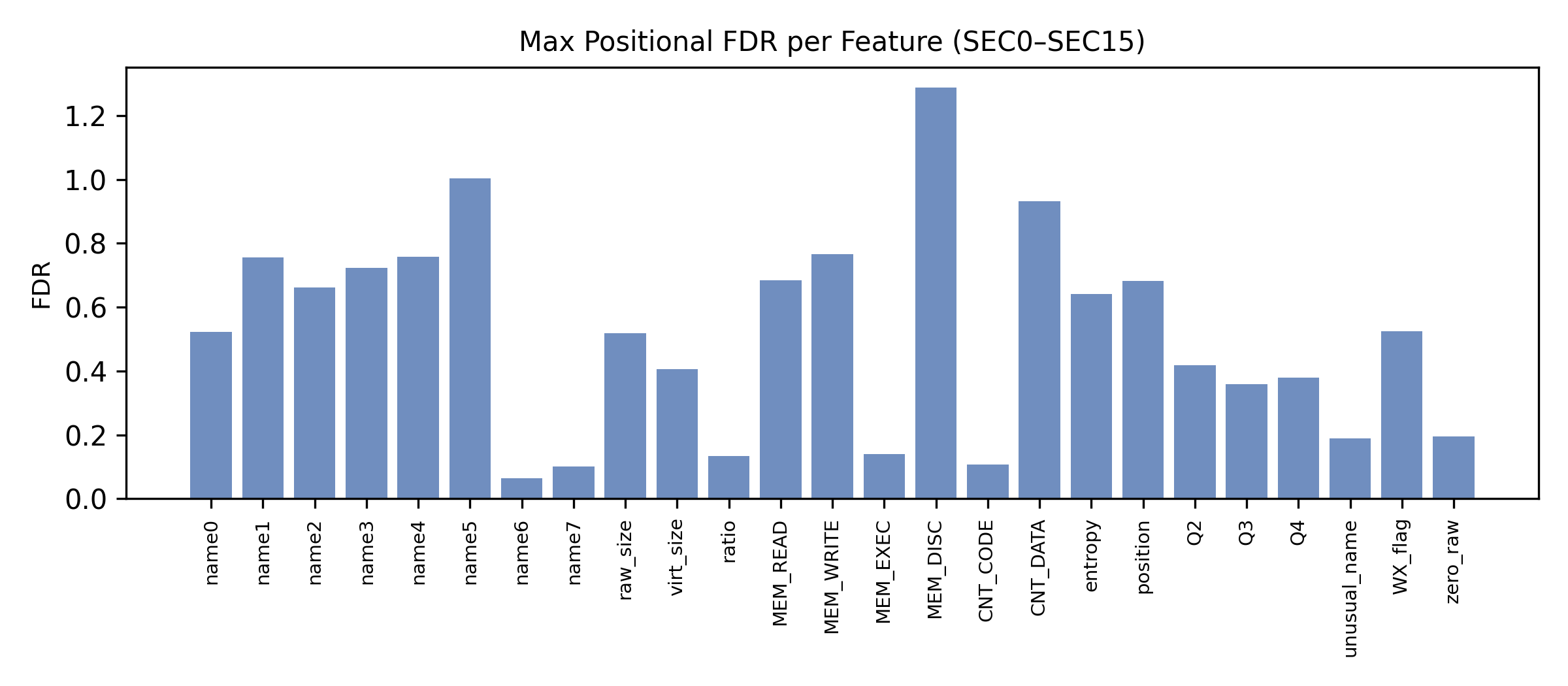}
  \caption{Maximum positional FDR per feature dimension (best of SEC0--SEC15).}
  \label{fig:fdr-a}
\end{subfigure}

\vspace{4pt}

\begin{subfigure}{0.49\columnwidth}
  \centering
  \includegraphics[width=\linewidth]{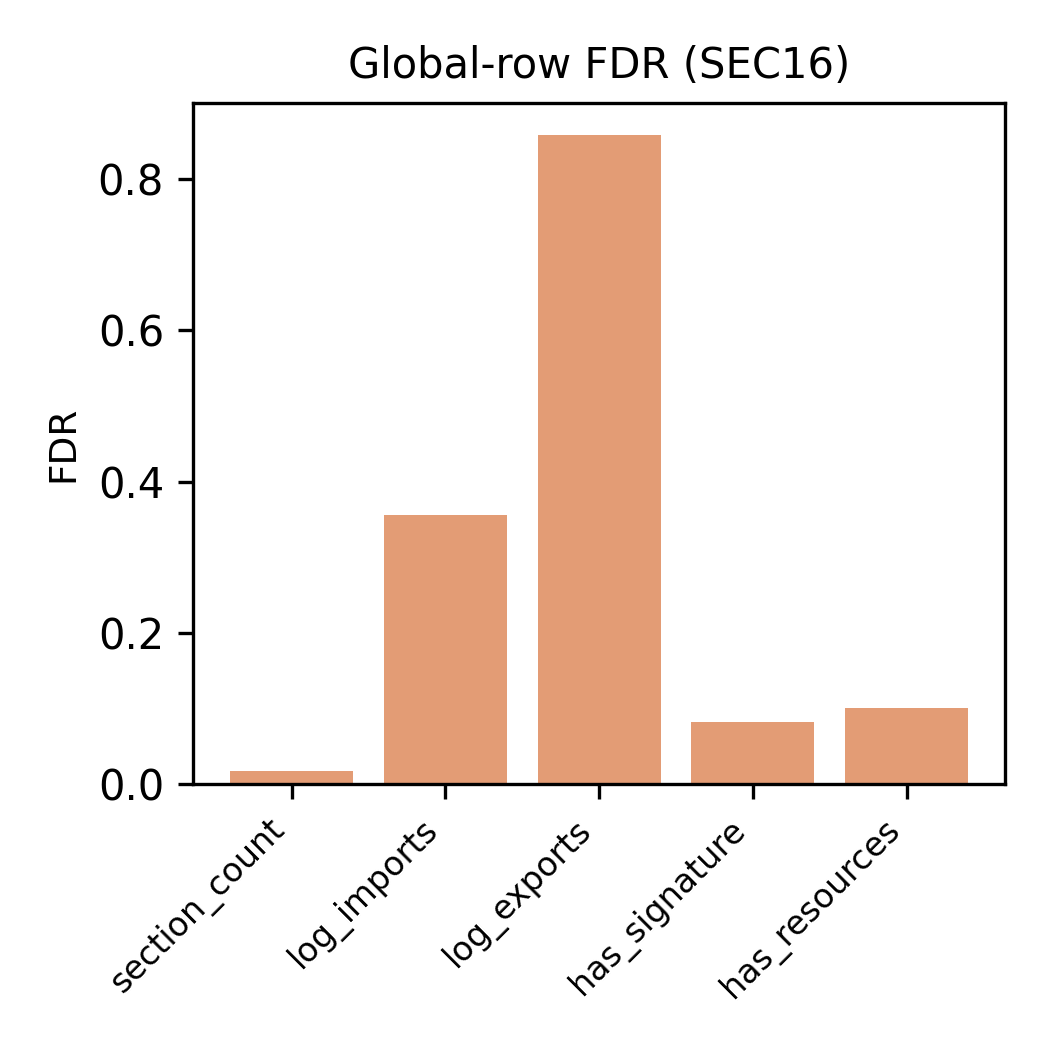}
  \caption{Global-row FDR (5 populated slots).}
  \label{fig:fdr-b}
\end{subfigure}
\hfill
\begin{subfigure}{0.49\columnwidth}
  \centering
  \includegraphics[width=\linewidth]{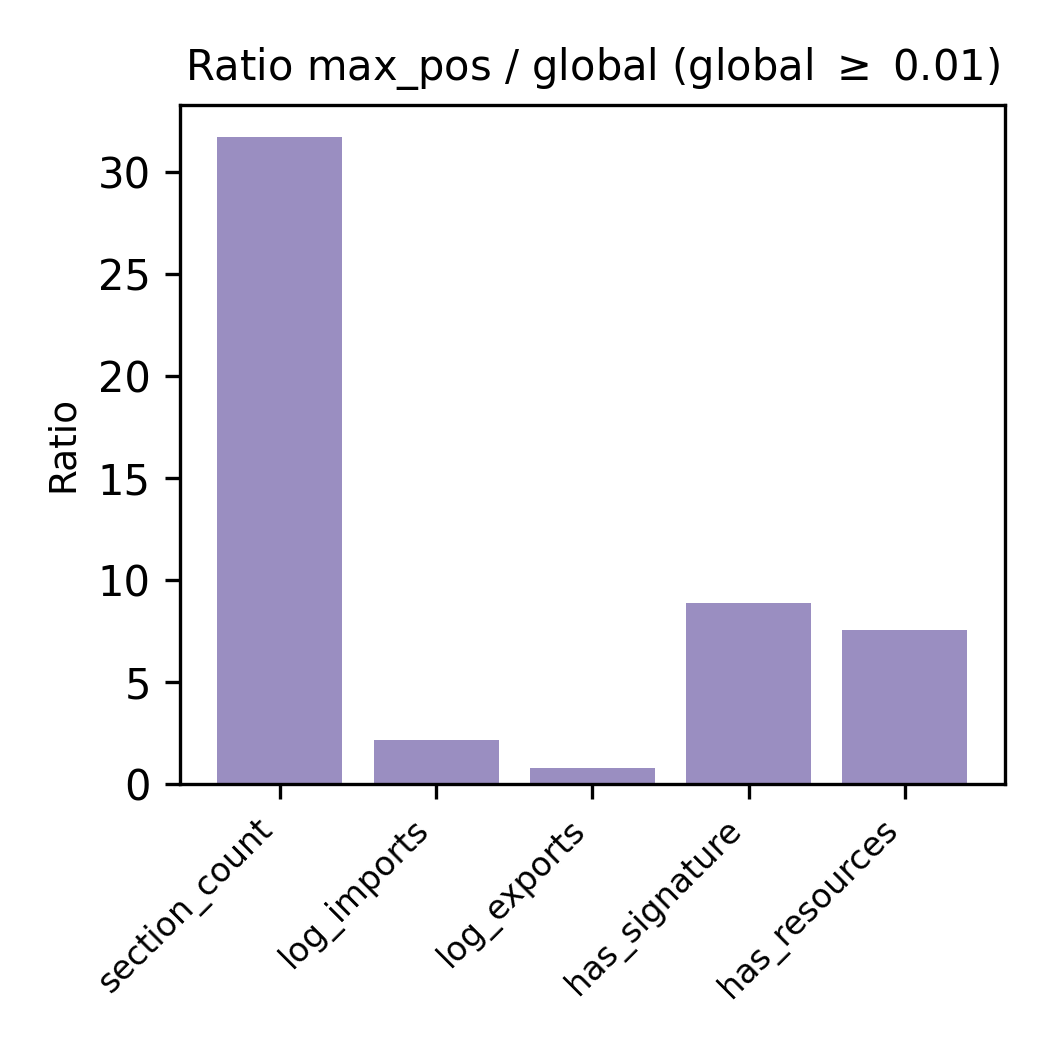}
  \caption{Positional-to-global FDR ratio.}
  \label{fig:fdr-c}
\end{subfigure}
\caption{Per-feature FDR comparison on the 49{,}204-sample family-filtered corpus. (a) Maximum positional FDR per feature dimension (best of SEC0--SEC15). (b) FDR of each feature dimension in the PRISM global summary row (row index 16); the global row populates 5 of 25 slots with file-level descriptors, the remaining 20 zero by construction. (c) Ratio of positional to global FDR for the 5 populated global slots. The maximum positional FDR (1.287, MEM\_DISC@SEC5) is $1.50\times$ the maximum global-row FDR (0.858, \texttt{log\_exports}).}
\label{fig:fdr-comparison}
\end{figure}

The highest-discriminating individual (section, feature) pair is \texttt{MEM\_DISC@SEC5} (FDR = 1.287): the MEM\_DISCARDABLE memory permission flag at section position 5. This reflects systematic differences between benign and malware in how discardable sections (typically \texttt{.rsrc} or \texttt{.reloc} content marked for discard after load) are positioned within the file: standard compiler toolchains place \texttt{.reloc} at consistent positions, while malware compilers and packers often produce different layouts. The top five (section, feature) pairs by FDR are \texttt{MEM\_DISC@SEC5} (1.287), \texttt{name5@SEC3} (1.004), \texttt{CNT\_DATA@SEC4} (0.931), \texttt{log\_exports@GLOBAL} (0.858), and \texttt{CNT\_DATA@SEC5} (0.853); Table~\ref{tab:separability} summarises the per-section versus global-row maxima for both FDR and MI, and Figure~\ref{fig:fdr-comparison} shows the per-feature comparison. The concentration of discriminative power in sections 1 through 5 --- with sections 7 and beyond contributing negligible signal --- is the most robust qualitative finding of the FDR analysis.

\textit{Interpretability of \texttt{name*} dimensions.} Several high-ranking cells (e.g., \texttt{name5@SEC3}, \texttt{name1@SEC*}) are individual bits of the 8-dimensional hashed section-name projection rather than directly nameable PE fields. These dimensions should be read collectively, as evidence that the \emph{distribution of section names at a given position} differs systematically between classes (a known packer/compiler signature), not as individually interpretable features. By contrast, \texttt{log\_exports@GLOBAL}, along with the permission flag (\texttt{MEM\_DISC}), size (\texttt{raw\_size}, \texttt{virt\_size}), and entropy cells that dominate the remaining rankings, map directly onto semantically meaningful PE attributes.

\subsection{Mutual Information}
\label{sec:mi}

Mutual Information (MI) measures the statistical dependence between a feature and the class label, capturing non-linear relationships that the linear FDR may miss. We estimate MI using the k-nearest-neighbour estimator of Kraskov et al.~\cite{kraskov_estimating_2004} with $k = 5$, as implemented in scikit-learn (\texttt{mutual\_info\_classif}, \texttt{random\_state=42}, \texttt{n\_neighbors=5}). We note that 9 of our 25 features are binary (the six permission flags READ, WRITE, EXEC, DISC, CODE, DATA, and the three anomaly flags), for which the Kraskov--St\"{o}gbauer--Grassberger estimator may exhibit known bias; mixed discrete--continuous MI estimators would provide refined values for those features and are listed as a methodological refinement in Section~\ref{sec:future}. MI is computed independently for each feature--position pair (section index $i$, feature $f$), following the same positional granularity used for FDR in Section~\ref{sec:fdr}.

\begin{figure}[t]
\centering
\includegraphics[width=\columnwidth]{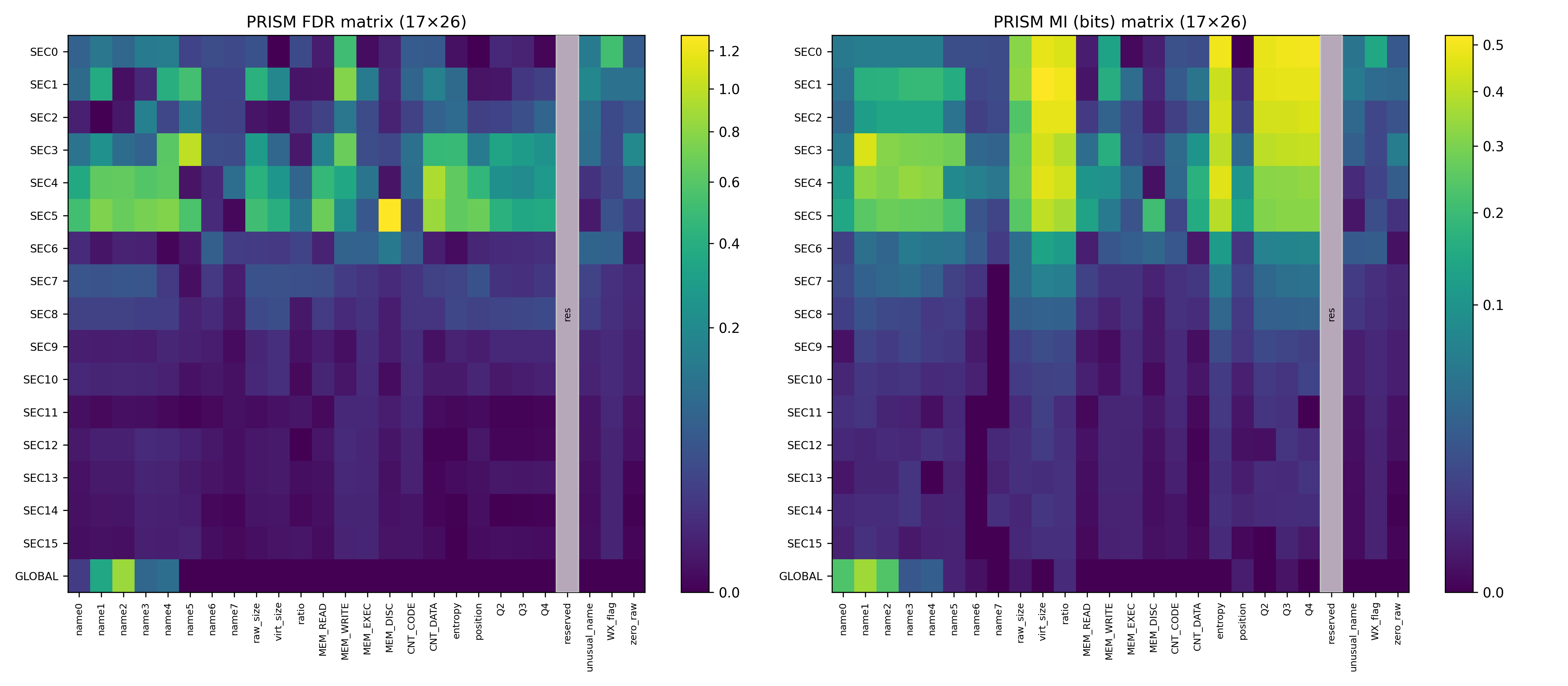}
\caption{FDR (left) and MI (right) heatmaps over the full $(17 \times 25)$ feature space on the 49{,}204-sample family-filtered corpus. Both heatmaps show discriminative power concentrated in sections SEC0--SEC5, with sections SEC7--SEC15 contributing negligibly. The FDR heatmap highlights MEM\_DISC@SEC5 (top value 1.287) and the cluster around sections 3--5; the MI heatmap highlights size and entropy-quartile features (virt\_size, ratio, Q2--Q4) concentrated at SEC0--SEC2. The `reserved' column (slot 22, see Section~\ref{sec:per-section-features}) is masked in grey.}
\label{fig:heatmaps}
\end{figure}

The results are visualised in Figure~\ref{fig:heatmaps}, which shows the FDR and MI values as heatmaps over the full $(\Nmax \times F)$ feature space. The top five (section, feature) pairs by MI are virt\_size@SEC1 (0.522 bits), Q4@SEC0 (0.499), entropy@SEC0 (0.495), Q3@SEC0 (0.494), and ratio@SEC1 (0.492). The MI ranking emphasises size, ratio, and entropy features at sections 0 and 1, in contrast to the FDR ranking which emphasises permission and name-encoding features at sections 3 to 5. Both rankings converge on the qualitative claim that discriminative information is concentrated in the first six section positions.

The maximum MI attained by any (section, feature) cell on the 
family-filtered corpus is 0.522 bits at virt\_size@SEC1, considerably 
higher than any populated dimension of the global summary row; the 
highest global-row MI is 0.357 bits (\texttt{log\_imports}), yielding 
a positional-to-global MI ratio of 1.46$\times$---consistent with the 
1.50$\times$ FDR ratio reported in Section~\ref{sec:fdr}.

\subsection{Inter-Cell Information Gain ($\dI$)}
\label{sec:delta-i}

Let $d = f^{(i)}_a - f^{(j)}_b$ denote the cross-cell feature 
difference between cells $(i, a)$ and $(j, b)$ of the matrix, 
where cells refer to (section index, feature index) positions. 
The inter-cell information gain is
\begin{equation}
\label{eq:delta-i}
\dI = I(d;\, Y) - \max\bigl\{I(f^{(i)}_a;\, Y),\; 
      I(f^{(j)}_b;\, Y)\bigr\}
\end{equation}
where $I(\cdot; Y)$ is the MI with the class label estimated 
under the same Kraskov $k$-NN procedure. The choice of the 
scalar difference $d = f^{(i)}_a - f^{(j)}_b$ makes $\dI$ a 
deliberately conservative, single-degree-of-freedom probe of 
inter-cell structure: it asks whether the \emph{relative} value 
of two cells is more informative than either cell alone, which 
is the specific relational quantity that a flat per-cell 
representation cannot encode. It is conservative precisely 
because it collapses the two cells onto a single additive 
contrast, capturing only the linear (difference) interaction 
between them and discarding any higher-order or non-additive 
dependence on the pair $(f^{(i)}_a, f^{(j)}_b)$ that the full 
joint distribution carries; consequently $\dI$ can only 
\emph{under}-report synergy, never overstate it. It is not the 
full joint mutual 
information $I(f^{(i)}_a, f^{(j)}_b;\, Y)$, which would upper-bound 
$\dI$ and which we leave to the refined estimation discussed in 
Section~\ref{sec:future}; the difference probe is preferred here 
because it is cheap to estimate over all 85{,}000 inter-section pairs and 
yields an interpretable lower bound on the synergy present in the 
lattice. Consistent with the inter-section motivation of PRISM, we 
restrict $\dI$ to pairs of cells lying in \emph{different} section 
rows (including the global summary row as one ``section''): with 
$\Nmax+1 = 17$ rows and $F = 25$ effective features per row, this is 
$\binom{17}{2}\times 25^2 = 136 \times 625 = 85{,}000$ unordered 
inter-section cell pairs. Pairs of features within a single section 
are excluded by construction, as the cross-section relationship is the 
quantity a flat 1D aggregation destroys. Of 
these, 62{,}950 pairs (74.1\%) exhibit $\dI > 0$ bits, 12{,}854 
pairs (15.1\%) exhibit $\dI > 0.01$ bits, and 2{,}215 pairs 
(2.6\%) exhibit $\dI > 0.05$ bits. The maximum $\dI = 0.205$ 
bits is attained by the pair (\texttt{raw\_size@SEC2}, 
\texttt{name5@SEC3}); the mean $\dI$ across pairs above the 
$0.01$ threshold is $0.033$ bits.

\begin{figure}[t]
\centering
\includegraphics[width=\columnwidth]{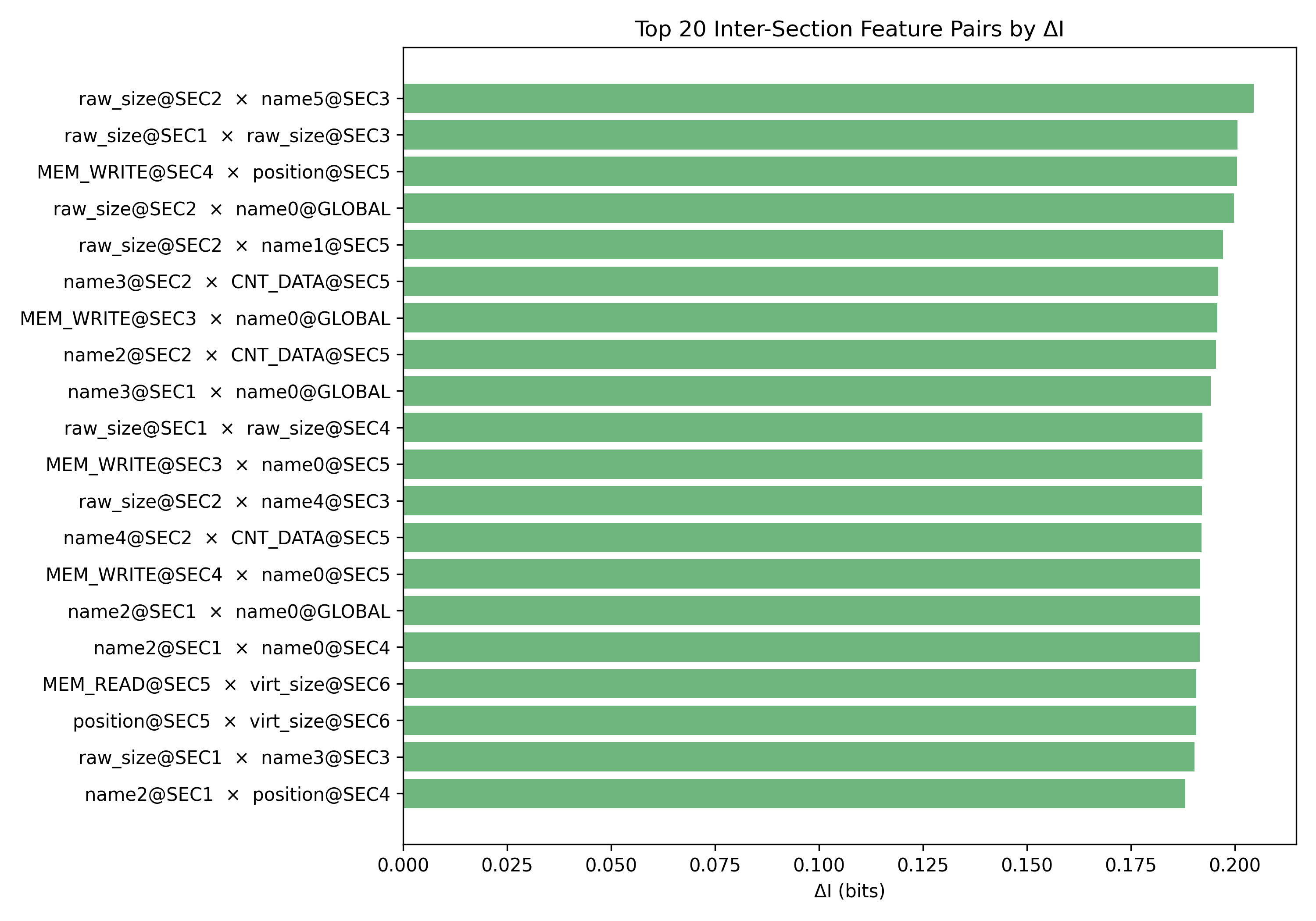}
\caption{Top 20 inter-section feature pairs by $\dI$ on the 49{,}204-sample family-filtered corpus. Each bar represents a pair of cells $(\text{section}_a, \text{feature}_a) \times (\text{section}_b, \text{feature}_b)$ whose joint MI exceeds the best individual MI by the indicated margin. The top pair (raw\_size@SEC2, name5@SEC3) contributes $\dI = 0.205$ bits.}
\label{fig:delta-i}
\end{figure}

The top 20 pairs by $\dI$ are shown in Figure~\ref{fig:delta-i}. Several patterns are visible: (i) cross-section pairs involving raw\_size at one section and a name-encoding feature at a different section appear repeatedly in the top 20 (e.g., raw\_size@SEC2 $\times$ name5@SEC3, raw\_size@SEC1 $\times$ raw\_size@SEC3, raw\_size@SEC2 $\times$ name1@SEC5), suggesting that the joint structure of size progression and naming across sections carries information unavailable from any individual cell; (ii) pairs involving the global summary row appear (e.g., raw\_size@SEC2 $\times$ name0@GLOBAL, name3@SEC1 $\times$ name0@GLOBAL), indicating that file-level descriptors and positional features carry partially complementary signal; (iii) the $\dI$ signal extends to non-adjacent section pairs (e.g., raw\_size@SEC1 $\times$ raw\_size@SEC4, MEM\_READ@SEC5 $\times$ virt\_size@SEC6), confirming that limiting analysis to adjacent transitions would miss substantial structural information.

The systematic presence of $\dI > 0.01$ in 15.1\% of all inter-section (section, feature) cell pairs constitutes the central representational finding of the separability analysis: discriminative information is widely distributed across the (section $\times$ feature) lattice in a form that requires \emph{joint} observation of two cells to be detected. Because a global aggregation collapses each feature across sections into a single file-level statistic, this position-dependent relational signal is not recoverable from the EMBER-style global vector --- although, as the saturation analysis of Section~\ref{sec:saturation} shows, recovering it confers no advantage on the \emph{binary} task once a position-free aggregate of the same features already saturates the metric.

\subsection{Entropy Profile by Section Position}
\label{sec:entropy-profile}

\begin{figure}[t]
\centering
\includegraphics[width=\columnwidth]{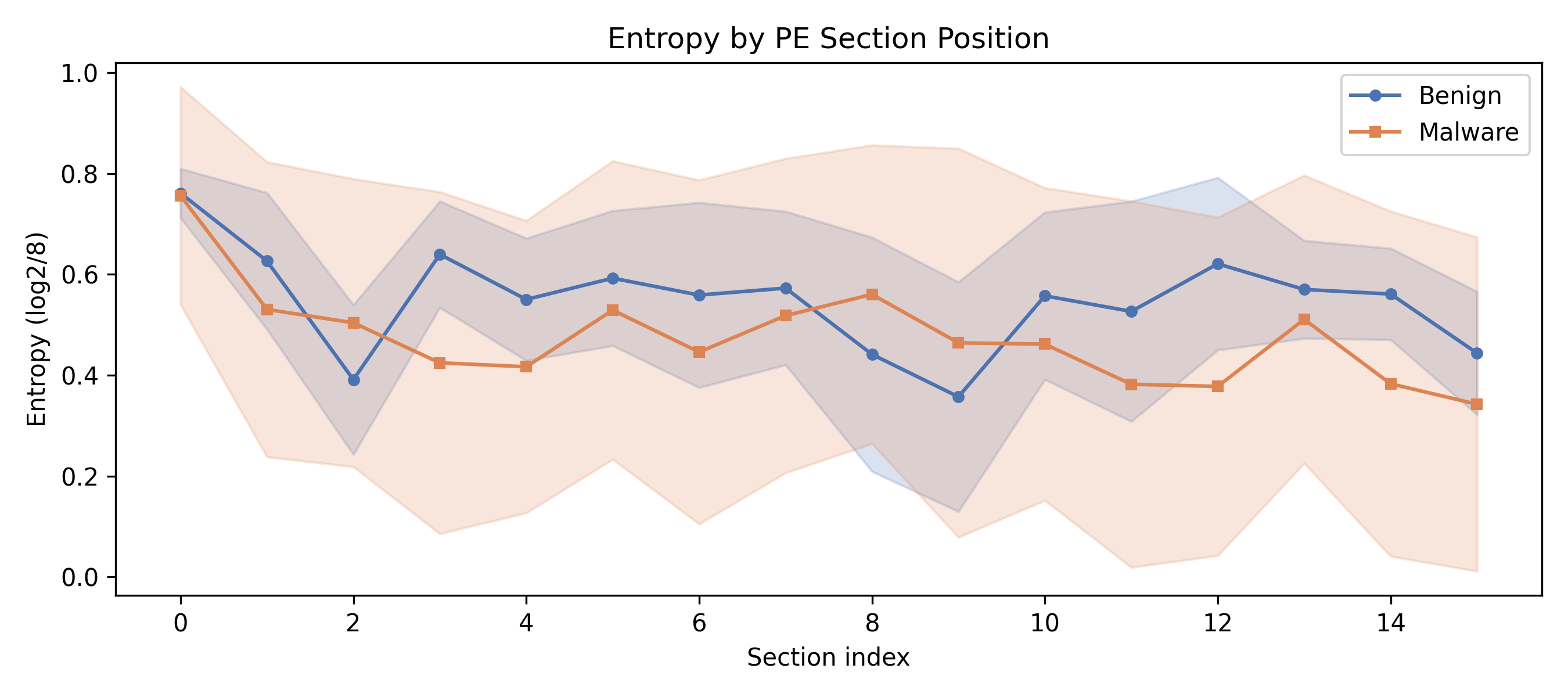}
\caption{Entropy profile by PE section position on the 49{,}204-sample family-filtered corpus. Both classes start near the same level at section 0 (benign $\approx 0.76$, malware $\approx 0.75$). Beyond section 0 the two profiles diverge, with benign exhibiting greater dispersion across sections (consistent with the more heterogeneous compiler and toolchain mix in the SOREL benign distribution) and malware showing a flatter, slightly lower mean profile. Shaded bands are $\pm 1$ standard deviation.}
\label{fig:entropy-profile}
\end{figure}

The entropy-by-position profile (Figure~\ref{fig:entropy-profile}) shows that both benign and malware classes start at near-equal entropy at section 0 (benign $\approx 0.76$, malware $\approx 0.75$), with the two profiles separating in subsequent sections in patterns consistent with the structural distinctions documented by FDR and MI. The present profile, computed on SOREL-derived benigns from heterogeneous compilers, exhibits a dispersed structure that is robust to compiler choice.

\textbf{Statistical test of per-position separation.} Visual inspection of Figure~\ref{fig:entropy-profile} suggests modest divergence between the benign and malware entropy distributions beyond section~0, but the heterogeneous SOREL-derived benign mix produces wide $\pm 1\sigma$ bands and makes the magnitude of any underlying distributional difference difficult to judge by eye. To place the visual observation on a quantitative footing without overclaiming, we apply a two-sample Kolmogorov--Smirnov (KS) test at each section position $i \in \{0, 1, \ldots, 15\}$, comparing the per-sample entropy distributions of the benign and malware populations on the full 49{,}204-sample family-filtered corpus. The KS statistic is non-parametric, makes no assumption about distribution shape, and detects any departure between the two empirical CDFs (location, scale, or higher moments). We report the per-position KS statistic $D_i$ and the two-sided $p$-value, applying Holm--Bonferroni correction across the 16 simultaneous tests to control the family-wise error rate at $\alpha = 0.05$. Active-row counts per position decline with section index (from $n_\text{benign} = 29{,}467$, $n_\text{malware} = 19{,}737$ at SEC0 to $n_\text{benign} = 187$, $n_\text{malware} = 165$ at SEC15), reflecting the section-count distribution of Figure~\ref{fig:section-distribution}.

The null hypothesis of distributional equality is rejected at every section position after Holm--Bonferroni correction (all corrected $p < 10^{-16}$; corrected $p$-values at the most populated positions SEC0--SEC5 underflow IEEE double precision). The KS statistic itself, which is bounded in $[0, 1]$ and serves as a non-parametric effect-size measure, ranges from $D = 0.262$ at SEC1 to $D = 0.597$ at SEC12, with a mean of $D = 0.408$ across the 16 positions (Table~\ref{tab:ks-results}). The largest separations occur at SEC12 ($D = 0.597$), SEC0 ($D = 0.485$), SEC15 ($D = 0.454$), and SEC14 ($D = 0.454$). These effect sizes are consistent with the qualitative reading of Figure~\ref{fig:entropy-profile}: a real but moderate distributional shift between classes that is not visually dramatic in mean-and-band plots but is highly statistically robust given the large per-position sample sizes available in the family-filtered corpus. The result formally confirms that the entropy-by-position structure carries discriminative content at \emph{every} section position, consistent with the broader FDR and MI patterns documented in Sections~\ref{sec:fdr}--\ref{sec:mi}.

\begin{table}[h]
\centering
\caption{Two-sample Kolmogorov--Smirnov test of benign vs.~malware Shannon entropy distributions at each section position on the 49{,}204-sample family-filtered corpus. Holm--Bonferroni correction is applied over the 16 simultaneous tests at $\alpha = 0.05$. $D$ is the KS statistic (bounded in $[0,1]$, larger = greater distributional separation). All 16 positions reject $H_0$ at the corrected level.}
\label{tab:ks-results}
\setlength{\tabcolsep}{4pt}
\renewcommand{\arraystretch}{1.05}
\footnotesize
\begin{tabular}{@{}c r r c c@{}}
\toprule
\textbf{Pos} & $\mathbf{n_\text{ben}}$ & $\mathbf{n_\text{mal}}$ & $\mathbf{D}$ & $\mathbf{p_\text{Holm}}$ \\
\midrule
SEC0  & 29{,}467 & 19{,}737 & 0.485 & $< 10^{-300}$ \\
SEC1  & 29{,}467 & 19{,}720 & 0.262 & $< 10^{-300}$ \\
SEC2  & 29{,}381 & 19{,}203 & 0.391 & $< 10^{-300}$ \\
SEC3  & 26{,}354 & 13{,}474 & 0.412 & $< 10^{-300}$ \\
SEC4  & 26{,}322 &  9{,}758 & 0.441 & $< 10^{-300}$ \\
SEC5  & 24{,}696 &  6{,}754 & 0.319 & $< 10^{-300}$ \\
SEC6  &  6{,}498 &  5{,}203 & 0.443 & $5.3 \times 10^{-321}$ \\
SEC7  &  1{,}993 &  3{,}122 & 0.350 & $1.2 \times 10^{-131}$ \\
SEC8  &  1{,}120 &  1{,}775 & 0.348 & $2.0 \times 10^{-73}$ \\
SEC9  &     769  &     742  & 0.400 & $1.3 \times 10^{-53}$ \\
SEC10 &     452  &     587  & 0.313 & $4.5 \times 10^{-22}$ \\
SEC11 &     329  &     263  & 0.422 & $3.5 \times 10^{-23}$ \\
SEC12 &     198  &     229  & 0.597 & $6.0 \times 10^{-35}$ \\
SEC13 &     194  &     191  & 0.439 & $6.6 \times 10^{-17}$ \\
SEC14 &     190  &     179  & 0.454 & $3.3 \times 10^{-17}$ \\
SEC15 &     187  &     165  & 0.454 & $7.2 \times 10^{-17}$ \\
\bottomrule
\end{tabular}
\end{table}

\section{Experimental Baselines}
\label{sec:baselines}

We define five model configurations (two of which share the PRISM flattened representation on different corpora) to assess the contribution of the PRISM representation under a strictly controlled experimental protocol. All models use LightGBM~\cite{ke2017lightgbm} as the classifier --- a gradient-boosted decision tree framework that has been the standard choice for tabular PE malware classification since the original EMBER paper~\cite{anderson2018ember} and that provides a strong and well-understood reference point. Table~\ref{tab:baselines-summary} summarises the five configurations at a glance; the subsections that follow describe each in detail, and Section~\ref{sec:exp-design} states the three hypotheses these models are designed to test.

\begin{table}[t]
\centering
\caption{Summary of the five model runs, grouped by the three contrasts of Section~\ref{sec:exp-design}. The primary cross-representation comparison ($\mprismsub$ vs.~$\membersub$) shares one EMBER-compatible sub-corpus and one split; the within-PRISM ablation ($\mprismpool$ vs.~$\mprismfull$) shares the full family-filtered corpus and one split; $\mprismtemp$ is the temporal probe. The three axes that vary are feature representation, corpus, and split protocol; the classifier (LightGBM) and hyperparameters are held fixed throughout.}
\label{tab:baselines-summary}
\setlength{\tabcolsep}{4pt}
\renewcommand{\arraystretch}{1.2}
\footnotesize
\begin{tabularx}{\columnwidth}{@{}l >{\raggedright\arraybackslash}X >{\raggedright\arraybackslash}X r@{}}
\toprule
\textbf{Model} & \textbf{Representation} & \textbf{Corpus / split} & \textbf{Dim} \\
\midrule
\multicolumn{4}{@{}p{\columnwidth}}{\textit{Contrast 1 --- PRISM vs.~EMBER (cross-representation, same samples):}} \\
\addlinespace
$\mprismsub$ & PRISM flattened & Sub-corpus 32{,}623 (SOREL\,+\,CAPE\,+\,MBZ); random 80/20, seed 42 & 425 \\
\addlinespace
$\membersub$ & EMBER 1D & Sub-corpus 32{,}623; identical split to $\mprismsub$ & 2{,}381 \\
\midrule
\multicolumn{4}{@{}p{\columnwidth}}{\textit{Contrast 2 --- within-PRISM ablation (positional vs.~position-discarded, same features):}} \\
\addlinespace
$\mprismpool$ & PRISM per-section features, mean-pooled (position discarded) & Full family-filtered (49{,}204); random 80/20, seed 42 & 25 \\
\addlinespace
$\mprismfull$ & PRISM flattened (per-section + global row) & Full family-filtered (49{,}204); random 80/20, seed 42 & 425 \\
\midrule
\multicolumn{4}{@{}p{\columnwidth}}{\textit{Contrast 3 --- single-class temporal probe (malware only):}} \\
\addlinespace
$\mprismtemp$ & PRISM flattened & BODMAS malware temporal (split 2020-07-25)\,+\,SOREL benign random; 36{,}557/9{,}141 & 425 \\
\bottomrule
\end{tabularx}
\end{table}

\begin{table*}[t]
\centering
\caption{Sample composition of each model by source. $\mprismpool$ and $\mprismfull$ share
the full family-filtered corpus; $\mprismtemp$ re-splits that corpus with a single-class
temporal cut on BODMAS malware (2020-07-25, the 80th-percentile first-seen
timestamp) while SOREL benigns are split at random; $\mprismsub$ and $\membersub$ use the
EMBER-compatible sub-corpus. VirusShare and MalwareBazaar~(legacy) are part of
the full deduplicated pool but are excluded from the family-filtered corpus, as
they lack verified family labels. Dashes denote sources not present in a model.}
\label{tab:baseline-composition}
\small
\setlength{\tabcolsep}{4pt}
\begin{tabularx}{\textwidth}{@{} >{\raggedright\arraybackslash}X r r r r r r @{}}
\toprule
\textbf{Model (split)} & \textbf{BODMAS} & \textbf{CAPE} & \textbf{MalwareBazaar} & \textbf{SOREL-20M} & \textbf{Malware} & \textbf{Total} \\
 & (malware) & (malware) & (modern, mal.) & (benign) & \textbf{total} & \\
\midrule
$\mprismpool$ / $\mprismfull$ (full family-filtered) & 16{,}231 & 3{,}130 & 376 & 29{,}467 & 19{,}737 & 49{,}204 \\
\midrule
\multicolumn{7}{@{}p{\textwidth}}{\textit{$\mprismtemp$ --- single-class temporal probe (BODMAS malware temporal; SOREL benign random):}} \\
\quad train & 12{,}984 & --- & --- & 23{,}573 & 12{,}984 & 36{,}557 \\
\quad test  & 3{,}247  & --- & --- & 5{,}894  & 3{,}247  & 9{,}141 \\
\midrule
\multicolumn{7}{@{}p{\textwidth}}{\textit{$\mprismsub$ / $\membersub$ --- EMBER-compatible sub-corpus, primary run (controlled cross-representation, 20 seeds; BODMAS not included):}} \\
\quad total & --- & 3{,}110 & 376 & 29{,}137 & 3{,}486 & 32{,}623 \\
\quad train (seed 42) & --- & --- & --- & 23{,}310 & 2{,}789 & 26{,}099 \\
\quad test (seed 42)  & --- & --- & --- & 5{,}827  & 697     & 6{,}524 \\
\midrule
\multicolumn{7}{@{}p{\textwidth}}{\textit{$\mprismsub$ / $\membersub$ (BODMAS-inclusive) --- confirmatory run (restored BODMAS binaries added; single seed):}} \\
\quad total & 16{,}202 & 3{,}110 & 376 & 29{,}137 & 19{,}688 & 48{,}825 \\
\bottomrule
\end{tabularx}

\vspace{2pt}
\footnotesize
\emph{Notes.} (i)~$\mprismsub$ and $\membersub$ share this identical EMBER-compatible
sub-corpus (SOREL, CAPE, MalwareBazaar) and split, differing only in representation
(PRISM 425-dim vs.~EMBER 2{,}381-dim); the confirmatory BODMAS-inclusive run
(Section~\ref{sec:b4}) adds 16{,}202 BODMAS samples with valid EMBER vectors for a
48{,}825-sample corpus. The full family-filtered corpus lists 16{,}231 BODMAS PRISM matrices; the 29-sample difference is BODMAS binaries whose EMBER vector could not be extracted, so they enter the PRISM-only full corpus but not the BODMAS-inclusive cross-representation run. (ii)~Sub-corpus malware $=3{,}110$ CAPE $+\ 376$ MalwareBazaar $=3{,}486$; of an
initial $32{,}973$ candidates, $350$ ($1.06\%$) failed EMBER extraction.
(iii)~The primary multi-seed study repeats the comparison over 20 stratified splits
(seeds 42--61) on this same 32{,}623-sample sub-corpus.
\end{table*}

\subsection{Common Protocol}

All baselines use the following LightGBM hyperparameters: 1{,}000 max estimators, learning rate 0.05, num\_leaves 63, min\_child\_samples 20, early stopping at 50 rounds without improvement, \texttt{random\_state=42}, and \texttt{scale\_pos\_weight} set to the class ratio of the relevant training set. Train/test splits are deterministic: 80/20 stratified per class, generated with \texttt{numpy.random.seed(42)}. Train/test row overlap after deduplication is verified to be zero for every baseline.

$\mprismpool$ and $\mprismfull$ operate on the full 49{,}204-sample family-filtered corpus. $\mprismtemp$ re-splits that same corpus with a single-class temporal cut: BODMAS malware is split by first-seen timestamp at the 80th percentile (2020-07-25), while the SOREL benigns, which carry no usable timestamp, are split at random (Section~\ref{sec:b3}). $\membersub$ (and its PRISM counterpart $\mprismsub$), introduced for the controlled cross-representation comparison, operate on the EMBER-compatible sub-corpus (Section~\ref{sec:b4}).

\subsection{Model $\mprismfull$ / $\mprismsub$: PRISM Flattened}
\label{sec:b2}

$\mprismfull$ (and its sub-corpus variant $\mprismsub$) is a LightGBM classifier trained on PRISM matrices flattened to a $17 \times 25 = 425$-dimensional vector by concatenating all section rows in order. This flattening preserves the positional information --- the features of SEC0 occupy dimensions 0--24, SEC1 occupies dimensions 25--49, and so on --- while making the representation compatible with standard tabular classifiers that expect a fixed-length input vector. $\mprismfull$ operates on the full 49{,}204-sample family-filtered corpus; the sub-corpus variant $\mprismsub$ uses the same representation on the 32{,}623-sample EMBER-compatible subset (Section~\ref{sec:b4}). The 80/20 split yields 39{,}363 training samples (15{,}790 malware, 23{,}573 benign) and 9{,}841 test samples (3{,}947 malware, 5{,}894 benign); \texttt{scale\_pos\_weight} = 1.493.

\subsection{Model $\mprismpool$: Position-Discarded Mean Aggregate}
\label{sec:b1-proxy}

To isolate the specific contribution of per-section \emph{position}, we train a LightGBM model on a position-discarded aggregate of each PRISM matrix: the column-wise mean of the section rows, collapsing the $(\Nmax \times F)$ per-section block to a single 25-dimensional vector with the same feature semantics as $\mprismfull$ but no positional structure. This mean-pooled vector retains the marginal distribution of every per-section feature (mean name-hash occupancy, mean size, mean entropy, mean quartiles, mean permission rates) while discarding which section each value came from. $\mprismpool$ is trained and evaluated on the identical full family-filtered corpus and split as $\mprismfull$, so any performance difference between $\mprismpool$ (25-dim, position-free) and $\mprismfull$ (425-dim, positional) is attributable exclusively to the per-section positional structure that flattening preserves and mean-pooling destroys.

\subsection{Model $\mprismtemp$: PRISM under a Single-Class Temporal Probe (BODMAS malware)}
\label{sec:b3}

$\mprismtemp$ applies the \emph{same} PRISM 425-dimensional classifier as $\mprismfull$, but replaces the random split with a temporal one on the BODMAS malware, to probe whether the saturation seen under random splitting is inflated by leakage between near-duplicate samples collected close in time. BODMAS malware is ordered by first-seen timestamp and cut at the $80$th percentile (2020-07-25): malware before that date trains, malware on or after it tests (Figure~\ref{fig:bodmas-temporal}). We are explicit about an important limitation: the temporal split applies \emph{only to the malware class}. The benign class is SOREL-20M, which carries no usable first-seen timestamp, so benigns are split at random (80/20). $\mprismtemp$ is therefore a single-class temporal check, not a fully temporal protocol; the exact per-class composition is train $= 12{,}984$ malware (temporal) $+\ 23{,}573$ benign (random) $= 36{,}557$, test $= 3{,}247$ malware (temporal) $+\ 5{,}894$ benign (random) $= 9{,}141$. We report the outcome of this probe, and discuss the extent to which it supports the absence of temporal-window leakage, in Section~\ref{sec:saturation}.

\subsection{Model $\membersub$: Controlled Cross-Representation 
Benchmark (EMBER 1D vs.~PRISM)}
\label{sec:b4}

$\membersub$ is the EMBER arm of the controlled cross-representation experiment.
To compare the PRISM 2D representation against the EMBER 1D
representation under maximally fair conditions---identical
samples, identical train/test split, identical classifier
hyperparameters---both representations must be extracted from
the same set of raw PE binaries. All four malware sources in our
corpus provide binaries suitable for EMBER extraction. SOREL-20M,
CAPE, and MalwareBazaar provide them directly; BODMAS provides
\emph{disarmed} binaries (PE \texttt{Machine} and
\texttt{Subsystem} header fields zeroed), which we restore with a
small utility that rewrites those two fields
(\texttt{Machine}\,$\to$\,\texttt{0x014c}, \texttt{Subsystem}\,$\to$\,2)
before extraction. We verified that restoration is necessary and
sufficient for EMBER: on the disarmed binaries, $47$ of the
$2{,}381$ EMBER features deviate from their restored values --- $3$
in the \texttt{general} group, $5$ in \texttt{section}, and $39$
in \texttt{imports} (with \texttt{Machine}\,=\,0, LIEF fails to
parse the import table, corrupting the \texttt{imports\_4} bucket
that carries the highest gain), while the byte-histogram and
byte-entropy groups are unaffected; on the restored binaries all
$47$ recover their correct values. EMBER extraction requires both a binary (to restore) and a paired PRISM matrix; of the $57{,}133$ BODMAS PRISM matrices, $75$ lack a binary on disk, leaving $57{,}058$ EMBER candidates. Extraction on these restored BODMAS binaries succeeds for $57{,}033$ of $57{,}058$ ($99.96\%$; the $25$ failures are unrelated LIEF parse errors).

We report this comparison on two corpora. The \emph{primary} comparison is a
20-seed robustness study on the $32{,}623$-sample EMBER-compatible
sub-corpus drawn from SOREL-20M, CAPE, and MalwareBazaar (this is
the sub-corpus for which an EMBER checkpoint was first available);
of an initial $32{,}973$ candidates, $350$ ($1.06\%$) failed EMBER
extraction under LIEF~0.14 due to an unpatched compatibility
gap (\texttt{lief.not\_found} attribute),\footnote{\label{fn:lief}%
The EMBER feature extractor used here was originally developed
for LIEF~0.9; this work uses LIEF~0.14. Two monkey-patches were
applied to enable extraction: (a)~NumPy type alias patch
(\texttt{np.int}/\texttt{np.float}/\texttt{np.bool}$\,\to\,$%
corresponding Python builtins, addressing deprecations in
NumPy~1.24+); and (b)~LIEF exception class patch
(\texttt{lief.bad\_format}/\texttt{bad\_file}/\texttt{pe\_error}%
\,$\to$\,bare \texttt{Exception}, addressing reorganisation
between LIEF~0.9 and 0.14). A residual gap
(\texttt{lief.not\_found}) caused the 1.06\% extraction failure
rate. Feature values may differ slightly between LIEF versions;
these differences are negligible at the saturation regime
documented in Section~\ref{sec:results}.}
leaving $32{,}623$ validated samples ($29{,}137$ SOREL benign;
$3{,}486$ malware $=3{,}110$ CAPE $+\ 376$ MalwareBazaar). The
\emph{confirmatory} comparison (which we label \emph{Path~A}) adds the restored BODMAS samples,
giving the full $48{,}825$-sample EMBER-valid family-filtered
corpus ($16{,}202$ BODMAS with valid EMBER vectors join the
$32{,}623$, minus residual failures). Throughout the paper, ``Path~A'' refers to this BODMAS-inclusive confirmatory run, in which the disarmed BODMAS binaries are restored and EMBER-extracted (Section~\ref{sec:b4}) so that BODMAS, previously excluded for lack of usable binaries, can join the controlled comparison.

\subsection{Evaluation Protocol and Metrics}
\label{sec:eval-protocol}

For the primary comparison both representations are trained and
evaluated under $20$ independent stratified $80/20$ splits
(seeds $42$--$61$), with LightGBM forced into deterministic mode
(\texttt{num\_threads}=1, \texttt{force\_row\_wise}=true,
\texttt{deterministic}=true) and \texttt{scale\_pos\_weight} set
per split to the training malware:benign ratio. For each split we
record TPR@FPR=0.1\%, AUC-ROC, and a paired McNemar test of the
two models' hard predictions on the shared test set. Multiple
seeds are essential here: at FPR=0.1\% the operating threshold on
a $\sim$5{,}800-benign test set is fixed by only a handful of
false positives, so a single split yields a TPR@0.1\% estimate
with a standard error of several tenths of a percentage point ---
larger than the effect being measured. Reporting the distribution
of the paired difference across splits, rather than one split's
point estimate, is what makes the comparison reproducible.

For $\mprismpool$ and $\mprismfull$ on the full corpus, and for the single-seed confirmatory runs, we additionally compute 95\% bootstrap confidence intervals on TPR@FPR=0.1\% using 1{,}000 resamples of the test set with replacement. We caution that at FPR=0.1\% the metric resolution is approximately one false positive per 1{,}000 benigns, so a single split's bootstrap interval is wide and an apparent overlap between two single-split intervals is a weak, optimistic criterion for equivalence; the 20-seed paired analysis of Section~\ref{sec:results} supersedes it for the primary comparison.

\subsection{Experimental Design and Hypotheses}
\label{sec:exp-design}

The five baseline configurations are not evaluated as independent points on a leaderboard. They are designed to support three contrasts, each posing a distinct, testable question about \emph{binary} malware/benign classification. We state the hypotheses explicitly here, before the results, and we order them by centrality to the paper's claim: the primary cross-representation question first, the within-PRISM ablation second, and the temporal safeguard third. This ordering separates the experimental design from the interpretation that follows in Section~\ref{sec:saturation}.

\paragraph{Contrast 1 (primary): $\mprismsub$ vs.~$\membersub$ --- PRISM vs.~EMBER, sample-controlled.} The central question of this paper is how closely the PRISM representation, which preserves per-section structure at $5.6\times$ lower dimensionality, can approach the established EMBER representation on the binary detection task. We compare them on an identical EMBER-compatible sub-corpus over 20 independent splits, with identical LightGBM hyperparameters; the only manipulated variable is the feature vector itself (PRISM flattened, 425 dimensions, vs.~EMBER 1D, 2{,}381 dimensions). Path~A removes the previous binary-availability obstacle, so we additionally confirm the comparison on the full BODMAS-inclusive corpus. \textbf{Hypothesis 1: PRISM attains binary-detection performance close to EMBER's at a fraction of the dimensionality on the same samples; any residual gap is small and confined to the deep-FPR tail.}

\paragraph{Contrast 2 (within-PRISM ablation): $\mprismpool$ vs.~$\mprismfull$ --- positional vs.~position-discarded.} Having established how PRISM compares to EMBER, we ask what the per-section \emph{position} contributes \emph{within} PRISM. This contrast uses the full family-filtered corpus ($n=49{,}204$), same 80/20 random split, same classifier, same hyperparameters, and the same underlying per-section features. The only manipulated variable is whether position is retained: $\mprismpool$ mean-pools the section rows into a single 25-dimensional position-free vector, while $\mprismfull$ keeps the full flattened matrix (425 dimensions) with position encoded by row offset. Any performance difference is attributable exclusively to the per-section positional structure. \textbf{Hypothesis 2: retaining per-section position adds measurable binary-detection signal beyond a position-discarded mean aggregate of the same features.}

\paragraph{Contrast 3 (single-class temporal probe): $\mprismtemp$ --- PRISM under a malware-temporal split.} $\mprismtemp$ uses the PRISM representation (not EMBER) and does not enter a paired comparison with $\mprismsub$ or $\membersub$. Its role is exploratory: to probe whether the saturation observed in Contrast~1 under a random split is partly an artefact of leakage between near-duplicate BODMAS-era samples collected close in time. We split BODMAS malware temporally (training before 2020-07-25, testing on or after) while benigns, lacking timestamps, are split at random, and observe whether PRISM remains saturated. As detailed in Section~\ref{sec:b3}, the single-class nature of this split and the benign-source confound limit its evidential weight. \textbf{Hypothesis 3 (weak form): PRISM does not collapse under a malware-temporal split; we do not claim this establishes full temporal generalisation, which a timestamped benign source would be required to test.}

\paragraph{Anticipated objection: can a 25-dim position-free aggregate really match a 425-dim positional matrix and a 2{,}381-dimensional EMBER vector?} A naive reading of Contrast~2 might suggest that a negative result ($\mprismpool$ statistically indistinguishable from $\mprismfull$, and from $\membersub$) would imply the extra dimensions are uninformative. This is not the correct reading. $\mprismpool$ is not an arbitrary low-dimensional model: mean-pooling preserves the full marginal distribution of every per-section feature --- mean entropy, mean size, mean permission rates, mean section-name occupancy --- and discards only \emph{where} in the file each value occurs. The signal that saturates the binary task (a binary's aggregate import/entropy/size profile) survives this aggregation almost intact; only the positional refinement is lost. EMBER's own gain analysis on $\membersub$ confirms that binary detection concentrates in a few structural-identity features: a single import-hash bucket (\texttt{imports\_4}) accounts for gain 419{,}167, approximately $5.5\times$ the second-ranked feature. Saturation at TPR@FPR=0.1\% $\approx 0.99$ is therefore a property of \emph{the binary classification task under gradient-boosted tabular classifiers, on a family-filtered corpus in which signed benign software is well represented} --- not a property of the input dimensionality, and not evidence that position is uninformative in general. The separability analysis of Section~\ref{sec:separability} establishes that the per-section positional information PRISM preserves is real and statistically robust (peak FDR $1.50\times$ that of the global row; KS rejection of benign-vs-malware distributional equality at all 16 section positions; 12{,}854 inter-section pairs carrying $\Delta I > 0.01$ bits). That information is reserved for tasks with greater metric headroom --- family-level multi-class, 2D-aware architectures, adversarial robustness --- where the binary-saturation ceiling does not constrain measurement.

Section~\ref{sec:results} reports the empirical outcomes of all three contrasts, beginning with the primary PRISM-vs-EMBER comparison. Section~\ref{sec:saturation} interprets the joint pattern in terms of representational saturation under gradient-boosted tabular classifiers.

\section{Results}
\label{sec:results}

\subsection{Quantitative Results}

Table~\ref{tab:results} reports the five model runs. We present the results in the order of the contrasts defined in Section~\ref{sec:exp-design}: the primary cross-representation comparison first ($\mprismsub$ vs.~$\membersub$), the within-PRISM ablation second ($\mprismpool$ vs.~$\mprismfull$), and the single-class temporal probe last ($\mprismtemp$).

\begin{table*}[t]
\centering
\setlength{\tabcolsep}{4pt} 
\caption{Model results, grouped by the three contrasts of Section~\ref{sec:exp-design}. Contrast~1 ($\mprismsub$ vs.~$\membersub$) is reported as mean\,$\pm$\,std of TPR@FPR=0.1\% over 20 deterministic stratified splits (seeds 42--61) on the 32{,}623-sample sub-corpus, with the BODMAS-inclusive Path~A run shown as a single-seed confirmation. Contrast~2 ($\mprismpool$ vs.~$\mprismfull$) is the within-PRISM positional ablation on the full corpus. Contrast~3 ($\mprismtemp$) is the single-class temporal probe. EMBER holds a small but consistent advantage in TPR@FPR=0.1\% across all 20 splits ($\Delta=+0.85\pm0.62$ pp, Wilcoxon $p<10^{-4}$), while the two are operationally indistinguishable at threshold 0.5 (paired McNemar median $p=0.06$).}
\label{tab:results}
\small

\begin{tabularx}{\textwidth}{@{} X l c c c c c @{}}
\toprule
\textbf{Model} & 
\textbf{Corpus} & 
\textbf{\makecell{n\_train /\\ n\_test}} & 
\textbf{\makecell{Input\\ dim}} & 
\textbf{\makecell{AUC-\\ ROC}} & 
\textbf{\makecell{TPR @\\ FPR=0.1\%}} & 
\textbf{\makecell{TPR @\\ FPR=1\%}} \\
\midrule
\multicolumn{7}{@{}p{\textwidth}}{\textit{Contrast 1 --- PRISM vs.~EMBER cross-representation (same samples):}} \\
\textbf{\makecell[l]{$\mprismsub$\\ (20-seed)}} & \makecell[l]{32{,}623-sample\\ sub-corpus} & 26{,}099 / 6{,}524 & 425 & 0.9994 & \textbf{0.9887\,$\pm$\,0.0058} & 0.9957 \\
\addlinespace
\textbf{\makecell[l]{$\membersub$\\ (20-seed)}} & \makecell[l]{32{,}623-sample\\ sub-corpus} & 26{,}099 / 6{,}524 & 2{,}381 & 0.9999 & \textbf{0.9971\,$\pm$\,0.0019} & 1.0000 \\
\addlinespace
\multicolumn{7}{@{}p{\textwidth}}{\quad\textit{Confirmation on the full BODMAS-inclusive corpus (Path~A, single seed):}} \\
$\mprismsub$ / $\membersub$ & \makecell[l]{48{,}825-sample\\ full corpus} & 39{,}060 / 9{,}765 & 425 / 2{,}381 & 0.9997 / 1.0000 & 0.9975 / 0.9987 & 0.9992 / 1.0000 \\
\midrule
\multicolumn{7}{@{}p{\textwidth}}{\textit{Contrast 2 --- within-PRISM positional ablation (same features, full corpus):}} \\
$\mprismpool$ (mean-pooled) & Full family-filtered & 39{,}363 / 9{,}841 & 25 & 0.99980 & 0.9949 & 0.9987 \\
\addlinespace
$\mprismfull$ (flattened)  & Full family-filtered & 39{,}363 / 9{,}841 & 425 & 0.99980 & 0.9924 & 0.9987 \\
\midrule
\multicolumn{7}{@{}p{\textwidth}}{\textit{Contrast 3 --- single-class temporal probe (malware only):}} \\
$\mprismtemp$ (malware-temporal)  & \makecell[l]{BODMAS mal.\ temporal\\ + SOREL ben.\ random} & 36{,}557 / 9{,}141 & 425 & 0.99970 & 0.9969 & 0.9997 \\
\bottomrule
\end{tabularx}
\end{table*}

\paragraph{Primary result --- a small, consistent EMBER edge in the deep tail; operational parity (Contrast 1).} The controlled cross-representation comparison is the central result of the paper, and we report it over 20 independent deterministic splits rather than one, because at FPR=0.1\% a single split's TPR estimate is dominated by a handful of false positives. Across the 20 splits, $\mprismsub$ (425-dim) attains TPR@FPR=0.1\% $= 0.9887 \pm 0.0058$ and $\membersub$ (2{,}381-dim) attains $0.9971 \pm 0.0019$. EMBER is ahead in \emph{all 20} splits, by a mean of $\Delta = +0.85$ pp (range $+0.14$ to $+2.15$ pp); a Wilcoxon signed-rank test on the paired per-split differences rejects equality ($p < 10^{-4}$, paired $t$-test $p \approx 9\times10^{-6}$). This is a real, reproducible advantage, and we state it plainly: \emph{EMBER retains a small but systematic edge over PRISM in the deep-FPR tail.} At the same time, the advantage is confined to that tail: a paired McNemar test on the two models' hard predictions on the shared test set does not reject at the conventional level in most splits (median $p = 0.06$; significant in 9 of 20), so the two representations are \emph{operationally indistinguishable} at the decision threshold. The earlier single-seed reading (seed 42: $\Delta = +0.14$ pp with overlapping bootstrap CIs) corresponds to one of the splits most favourable to PRISM and should not be generalised --- a concrete illustration of why the multi-seed protocol is necessary. The practical takeaway is that PRISM recovers essentially all of EMBER's binary-detection performance, falling short only marginally and only in the extreme low-FPR regime, while using $5.6\times$ fewer input dimensions and preserving an interpretable per-section structure that EMBER discards.

\paragraph{Confirmation with BODMAS included (Path~A).} Lifting the BODMAS exclusion (Section~\ref{sec:b4}) and re-running the comparison on the full $48{,}825$-sample corpus reproduces the same ordering: $\mprismsub$ (BODMAS-inclusive) reaches AUC-ROC = 0.99965 and TPR@FPR=0.1\% = 0.9975, $\membersub$ (BODMAS-inclusive) reaches AUC-ROC = 0.99999 and TPR@FPR=0.1\% = 0.9987, a difference of $+0.127$ pp in the same direction. The BODMAS-inclusive corpus thus does not change the conclusion; we report it as a single-seed confirmation and note that a full multi-seed repetition on the expanded corpus is the natural next step.

At the operationally relevant decision threshold of 0.5, the two representations produce nearly identical confusion matrices. Table~\ref{tab:confusion} shows a representative split (seed 42) on the 6{,}524-sample sub-corpus test set (5{,}827 benign / 697 malware); the small false-negative difference shown there is within the per-split variation documented above and should not be read as a systematic cost-balanced advantage for either model.

\begin{table}[h]
\centering
\caption{Confusion matrices for $\mprismsub$ and $\membersub$ on the sub-corpus test set (6{,}524 samples) at decision threshold 0.5, for a single representative split (seed 42). Both commit identical false-positive counts (FP = 2); the four-sample false-negative difference is within the per-split variation reported in Section~\ref{sec:results} and is not a systematic effect.}
\label{tab:confusion}
\begin{tabular}{@{}lcccc@{}}
\toprule
\textbf{Model} & \textbf{TN} & \textbf{FP} & \textbf{FN} & \textbf{TP} \\
\midrule
$\mprismsub$ & 5{,}825 & 2 & 6  & 691 \\
$\membersub$ & 5{,}825 & 2 & 10 & 687 \\
\bottomrule
\end{tabular}
\end{table}

In this split both models commit exactly 2 false positives at threshold 0.5 (FPR = 0.034\%), with 6 false negatives for $\mprismsub$ and 10 for $\membersub$ out of 697 malware; across the 20 splits neither model holds a consistent advantage at threshold 0.5 (McNemar median $p = 0.06$), in contrast to the deep-tail metric where EMBER leads consistently. The two representations thus differ measurably only in the extreme low-FPR regime.

\paragraph{Within-PRISM ablation (Contrast 2).} Having characterised how PRISM compares to EMBER, we isolate the contribution of the per-section structure within PRISM. On the full family-filtered corpus, $\mprismpool$ (per-section features mean-pooled across sections, position discarded, 25-dim) reaches AUC-ROC = 0.99980 and TPR@FPR=0.1\% = 0.9949, while $\mprismfull$ (full flattened matrix, 425-dim, position retained) reaches AUC-ROC = 0.99980 and TPR@FPR=0.1\% = 0.9924. The two are statistically indistinguishable on the binary task; their bootstrap CIs ([0.9704, 0.9980] for $\mprismpool$ and [0.9578, 0.9962] for $\mprismfull$) overlap almost entirely. In binary classification at this saturation level, retaining per-section position adds no measurable detection signal over a position-free mean aggregate of the same features --- a result we interpret in Section~\ref{sec:saturation} and which must be read against the anticipated objection addressed in Section~\ref{sec:exp-design}: it reflects metric saturation, not absence of structural information.

\paragraph{Single-class temporal probe (Contrast 3).} $\mprismtemp$ applies the PRISM 425-dim classifier with a temporal split of the BODMAS malware (training before 2020-07-25, testing on or after) and a random split of the SOREL benigns. The result (AUC-ROC = 0.99970, TPR@FPR=0.1\% = 0.9969) shows that PRISM does not collapse under a malware-temporal split. We read this cautiously: because only the malware class is temporally separated and the benign source is, as the source probe below shows, trivially separable, $\mprismtemp$ is consistent with the absence of temporal-window leakage but does not establish full temporal generalisation. It is supporting evidence, not a decisive safeguard; a timestamped benign source would be needed for a complete test.

\subsection{Interpretation of the Saturation Regime}
\label{sec:saturation}

The empirical pattern observed in Section~\ref{sec:results} admits three complementary interpretations that map onto the three hypotheses stated in Section~\ref{sec:exp-design}.

\textbf{Observation 1 (answers Hypothesis~1): PRISM 425-dim nearly matches EMBER 2{,}381-dim, with a small consistent tail gap.} Over 20 splits, EMBER leads PRISM in TPR@FPR=0.1\% in every split, by a mean of $+0.85$ pp (Wilcoxon $p<10^{-4}$), yet the two are indistinguishable at the decision threshold (McNemar median $p=0.06$). PRISM therefore recovers nearly all of EMBER's binary-detection performance using $5.6\times$ fewer features while preserving an interpretable per-section structure that EMBER lacks, conceding only a marginal, tail-confined gap. This is primarily a representational-efficiency result, directly attributable to the design choice of preserving structural identity rather than aggregating into byte-histogram and string-bucket statistics that the gradient-boosted classifier must then implicitly reconstruct; the residual gap suggests EMBER's far larger feature set still captures a little additional deep-tail signal that the compact PRISM vector does not.

\textbf{Observation 2 (answers Hypothesis~2 negatively, in this regime): a position-discarded mean aggregate already saturates the binary task under tabular classifiers.} $\mprismpool$ (25-dim, per-section features mean-pooled across sections) achieves AUC = 0.99980 and TPR@FPR=0.1\% = 0.9949 on the full family-filtered corpus --- within the bootstrap confidence interval of the 425-dim $\mprismfull$ trained on the \emph{same} corpus and split (TPR@FPR=0.1\% = 0.9924, 95\% CI [0.9578, 0.9962]). Because $\mprismpool$ retains the same features as $\mprismfull$ and differs only in discarding section position, this isolates position as the manipulated variable: the mean per-section feature profile carries essentially all the binary discrimination available to a tabular classifier in this regime. Hypothesis~2 is therefore not supported under flat-vector gradient-boosted classification on this corpus: per-section positional structure does exist and is statistically richer (see Observation~3), but its additional signal cannot manifest while the position-free aggregate already saturates the metric. As emphasised in Section~\ref{sec:exp-design}, this is a statement about metric saturation in \emph{binary} classification, not about the absence of structural information.

\textbf{Observation 3 (reconciles Observations~1 and~2 with the separability analysis): separability content is real and present but not exploited by flat-vector tabular classifiers.} Section~\ref{sec:separability} documents that positional features carry substantially more per-feature discriminative content ($1.50\times$ higher maximum FDR than the global row's maximum) and that 12{,}854 inter-section pairs carry joint information beyond any individual cell. This structural information does not manifest in binary classification metrics at the saturation point because LightGBM operating on flattened features cannot exploit the 2D structure (it sees the matrix as an unordered bag of 425 scalars), and because binary classification at AUC $\approx 0.9999$ leaves no headroom for additional signal to manifest. The structural information becomes consequential under (i) 2D-aware architectures (CNN, ViT, GNN) that can natively exploit the matrix structure; (ii) fine-grained classification tasks (family-level multi-class with 684 classes provides substantially more headroom); (iii) adversarial and concept-drift evaluation, where representational richness affects robustness.

\textbf{Observation 4 (answers Hypothesis~3, weak form): no dominant temporal-window leakage artefact, but not a proof of temporal generalisation.} Under the single-class temporal split (Section~\ref{sec:b3}), $\mprismtemp$ reaches AUC-ROC = 0.99970 and TPR@FPR=0.1\% = 0.9969 on the 2020 test fold --- it does not collapse, which is consistent with the absence of leakage between near-duplicate samples collected close in time. We do not over-read this. Only the malware class is temporally separated; the benign class (SOREL) is split at random and, as the source probe above shows, benign provenance is itself trivially separable, so a model can sustain high apparent performance without genuinely generalising across the malware time axis. $\mprismtemp$ therefore \emph{supports}, but does not by itself \emph{establish}, that the PRISM--EMBER near-equivalence of Observation~1 is free of temporal leakage; a timestamped benign source would be required for a decisive test, which we identify as future work.

A qualitative convergence between the dominant features of the two representations supports the equivalence claim of Observation~1. The $\membersub$ model's top feature by gain is \texttt{imports\_4} (gain 419{,}167 --- approximately $5.5\times$ the second-ranked feature), an import-hash bucket. The $\mprismfull$ model's top three features by gain are \texttt{log\_exports@GLOBAL} (the global-row slot encoding log-scaled export count), \texttt{entropy@SEC0}, and \texttt{Q4@SEC1}. In both representations, structural identity features (encoding the binary's imports/exports profile and section-level entropy distribution) dominate the gain importance: the two representations capture the same underlying signal through different encodings.

\subsection{Limitations}

\textbf{LIEF version compatibility for $\membersub$.} The EMBER extractor was originally developed for LIEF 0.9 and is run here under LIEF 0.14 with two compatibility patches (footnote~\ref{fn:lief}). EMBER itself emits a warning about possible feature-value differences between LIEF versions. At the saturation regime documented above, these differences are expected to be negligible, but a fully matched reproduction under LIEF 0.9 would be the technically cleanest version of this experiment.

\textbf{Kraskov MI estimator bias for binary features.} The MI estimator used in Section~\ref{sec:mi} may produce biased estimates for the 9 binary features in the PRISM vector. A mixed discrete--continuous estimator (Ross 2014 or Gao et al.~2017) would refine those values; we have not re-run the analysis with that estimator in the present release.

\textbf{BODMAS disarming and restoration.} BODMAS binaries are distributed disarmed (\texttt{Machine} and \texttt{Subsystem} zeroed). We restore these two fields before EMBER extraction; on the disarmed form, 47 of 2{,}381 EMBER features deviate (chiefly 39 import features, because LIEF fails to parse the import table when \texttt{Machine}\,=\,0), all of which recover on the restored binaries. While restoration is faithful for the affected header and import features and leaves the byte-histogram and byte-entropy groups untouched, it is a reconstruction rather than the original vendor binary; a fully matched reproduction from never-disarmed BODMAS binaries would be the technically cleanest version of the BODMAS portion of the cross-representation benchmark. The multi-seed primary comparison (Section~\ref{sec:results}) is computed on the non-BODMAS sub-corpus and is unaffected by this caveat; the BODMAS-inclusive run is reported as confirmation only.

\textbf{Single-seed BODMAS-inclusive run.} The 48{,}825-sample BODMAS-inclusive comparison is reported for a single split, whereas the primary sub-corpus comparison uses 20 seeds. A multi-seed repetition on the expanded corpus is left to a revision; the single-seed expanded result is directionally consistent with the multi-seed sub-corpus result.

\textbf{Source provenance and disarming asymmetry (threat to validity for absolute metrics).} Benign and malware samples in this corpus are drawn from disjoint repositories: every benign sample originates from SOREL-20M, and no malware sample does. The disarming status is likewise asymmetric --- SOREL and BODMAS samples are distributed in disarmed form (subsystem and machine header fields zeroed), whereas CAPE and MalwareBazaar binaries are not; in the EMBER-compatible sub-corpus specifically, the benign class is entirely disarmed SOREL while the malware class (CAPE + MalwareBazaar) is not. A classifier can in principle exploit such provenance and preprocessing artefacts as a shortcut for the label, and we confirm this empirically: a LightGBM trained to predict the \emph{source} (SOREL vs.~non-SOREL) rather than the label achieves AUC = 0.99988, essentially identical to the malware detector's own AUC. Source and label are thus near-perfectly collinear in this corpus, and the detector cannot be distinguished from a provenance classifier on the available data. We consider this the most likely explanation for the very high \emph{absolute} operating point (AUC $\approx 0.9999$), which sits far above the EMBER2018 reference. The dominance of an import-hash bucket (\texttt{imports\_4}) and a header field (\texttt{header\_28}) in the $\membersub$ gain ranking is consistent with this concern. Crucially, this confound does \emph{not} undermine the two controlled comparisons that carry the paper's claims: in both $\mprismsub$-vs-$\membersub$ (cross-representation) and $\mprismpool$-vs-$\mprismfull$ (within-PRISM ablation), the compared models are trained and tested on the \emph{identical} samples, splits, and labels, so any provenance shortcut is shared equally between them and cancels in the comparison. The confound therefore inflates the shared ceiling but leaves the relative conclusions --- the near-equivalence of PRISM and EMBER, and the saturating role of the position-free per-section aggregate --- intact. It does, however, mean that the \emph{absolute} detection numbers in this paper must not be read as field performance estimates; a source-balanced corpus (benign and malware from a common feed) is required before any absolute detection claim is made, and we identify this as essential future work.

\textbf{Single deterministic split.} All baselines use one 80/20 split generated with \texttt{seed=42}. The bootstrap confidence intervals quantify test-set sampling variability but not split-selection variability. Repeated-split or stratified $k$-fold validation, and multiple seeds, would tighten the variance estimates; given the saturated regime we expect the qualitative conclusions to be stable, but we have not verified this across splits in the present release.

\section{Comparison with Related Datasets}
\label{sec:comparison}

\begin{table}[t]
\centering
\caption{Comparison of open PE malware datasets.}
\label{tab:dataset-comparison}
\small
\setlength{\tabcolsep}{5pt} 
\begin{tabularx}{\columnwidth}{@{} X c c c c @{}}
\toprule
\textbf{Dataset} & \textbf{\makecell{Unique\\ samples}} & \textbf{Year} & \textbf{\makecell{2D\\ repr.}} & \textbf{\makecell{2024--25\\ data}} \\
\midrule
EMBER~\cite{anderson2018ember}   & 1.1M  & 2018 & No  & No  \\
\addlinespace
BODMAS~\cite{yang2021bodmas}     & 134K  & 2021 & No  & No  \\
\addlinespace
SOREL-20M~\cite{harang2020sorel} & 20M   & 2020 & No  & No  \\
\addlinespace
EMBER2024~\cite{joyce2025ember2024} & 3.2M & 2025 & No  & Yes \\
\addlinespace
Multi-feature~\cite{yousuf2022multifeature} & 18K & 2022 & No & No \\
\midrule
\textbf{PRISM (ours)} & \textbf{83{,}633} & \textbf{2026} & \textbf{Yes} & \textbf{Yes} \\
\addlinespace
\raggedright \scriptsize (49{,}204 family-filtered) & & & & \\
\bottomrule
\end{tabularx}
\end{table}

PRISM is, to our knowledge, the first open PE malware dataset to provide a structured 2D matrix representation in which rows correspond to explicit PE sections in file order and columns correspond to named semantic feature dimensions, preserving section identity as a first-class addressable axis. While image-based representations~\cite{malik_matin_image_2025,hai_proposed_2023,yu_semantic_2025,zhao_mcpds_2026} preserve spatial structure implicitly through pixel arrangement, they do not expose section boundaries or named feature dimensions as addressable units, and therefore do not support per-section interpretability tools such as the FDR and MI heatmaps presented in Section~\ref{sec:separability}. PRISM also includes PE malware samples from 2024--2025 collected from MalwareBazaar alongside re-processed BODMAS data, addressing the temporal coverage gap identified in Section~\ref{sec:intro}.

\section{Conclusion}
\label{sec:conclusion}

We have introduced PRISM (PE Relational Inter-Section Matrix), an open dataset and 2D feature representation for static Windows PE malware detection. PRISM encodes each binary as a structured matrix in which rows correspond to PE sections in file order and columns correspond to named semantic feature dimensions, preserving the positional structure that flat 1D representations discard.

Our separability analysis on the 49{,}204-sample family-filtered corpus provides quantitative evidence that the per-section structural information preserved by PRISM is statistically detectable. The most discriminative individual feature (MEM\_DISC@SEC5) achieves a Fisher Discriminant Ratio of 1.287, $1.50\times$ higher than the maximum attained by any single dimension of the global summary row. Across the (section $\times$ feature) lattice, 12{,}854 inter-section pairs carry information gain $\dI > 0.01$ bits beyond either component alone, with the top pair (raw\_size@SEC2 $\times$ name5@SEC3) contributing $\dI = 0.205$ bits. These findings provide formal evidence that the 2D structure preserves discriminative information unavailable to 1D representations.

At the classifier level, the controlled cross-representation comparison ($\mprismsub$ vs.~$\membersub$ on an identical 32{,}623-sample sub-corpus over 20 deterministic splits) yields a small but consistent advantage for EMBER in the deep-tail metric: $\Delta$TPR@FPR=0.1\% = $+0.85 \pm 0.62$ pp, with EMBER ahead in all 20 splits (Wilcoxon $p < 10^{-4}$). At the decision threshold the two are operationally indistinguishable (paired McNemar median $p = 0.06$), and including BODMAS via restored binaries on the full 48{,}825-sample corpus reproduces the same ordering ($+0.127$ pp, single seed). \textbf{PRISM thus recovers nearly all of EMBER's binary-detection performance using $5.6\times$ fewer features (425 vs.~2{,}381) and a per-section interpretable structure, conceding only a marginal, tail-confined gap.} We are explicit that this near-equivalence is established in a saturated regime: the within-corpus comparison between $\mprismfull$ (425-dim, positional) and $\mprismpool$ (25-dim, position-discarded mean aggregate) is statistically indistinguishable, indicating that a position-free aggregate already saturates the binary detection task on this corpus under gradient-boosted decision trees. The high absolute operating point should be read with the provenance caveat of Section~\ref{sec:results} (a source-vs-label probe reaches AUC = 0.9999) firmly in mind: these are not field performance estimates. All baselines converge at AUC-ROC $\gtrsim 0.9997$ and TPR@FPR=0.1\% $\in [0.989, 0.997]$, establishing the saturation level for static binary classification on the family-filtered PRISM corpus and motivating the move to representations and tasks with genuine metric headroom.

The structural information preserved by PRISM --- documented quantitatively by the separability analysis --- is positioned to become consequential under architectures that can natively exploit the 2D matrix structure (convolutional networks, vision transformers, graph neural networks) and in fine-grained tasks (family classification, temporal drift, adversarial robustness) where binary metrics do not saturate. The integration of complementary dynamic features as a third tensor dimension is the natural next step in the research programme.

The PRISM dataset, the \texttt{prism-extract} extraction library, all trained models, the controlled cross-representation benchmark code, and the complete analysis pipeline are publicly available at \url{https://github.com/drjmsacristan/prism-dataset}, with the PRISM feature matrices and baseline models permanently archived and citable via Zenodo at \url{https://doi.org/10.5281/zenodo.20480349}.

\section{Future Work}
\label{sec:future}

Several directions extend naturally from the contributions of this paper. We list them in order of expected impact.

\textbf{2D-aware deep learning architectures.} The baselines presented in this paper deliberately treat the PRISM matrix as a flattened 1D vector to enable controlled comparison against EMBER and the global-row proxy under standard tabular classifiers. The natural next step is to evaluate architectures that exploit the 2D structure natively: convolutional networks with kernels over the (section, feature) plane; vision transformers with section-level attention modelling long-range cross-section dependencies; and graph neural networks that represent the section sequence as a directed graph with feature-level edges. The separability analysis of Section~\ref{sec:separability} documents the discriminative content these architectures could exploit; the present binary classification results suggest they will not produce gains on the binary task at the present saturation point but are positioned to produce gains on fine-grained tasks.

\textbf{Family-level classification.} With 684 distinct families in the family-filtered corpus and a heavy-tailed distribution (top-5 families cover 44.8\%, top-30 cover 73.6\%), the corpus supports multi-class evaluation. Family-level metrics will not saturate as readily as binary metrics; we expect the PRISM structural information to be more clearly advantageous at this granularity.

\textbf{3D extension with dynamic features.} The PRISM matrix is designed to accommodate a natural extension to three dimensions by incorporating dynamic analysis features --- API call sequences, memory access patterns, and network behaviour indicators --- as additional layers alongside the existing static per-section features. This extension is currently under active development with a target corpus of 30{,}000 malware and 30{,}000 benign samples under controlled sandbox execution. The cross-representation comparison ($\mprismsub$ vs.~$\membersub$) suggests that the most likely source of additional discriminative signal --- given that static representations have saturated on the binary task --- is the integration of complementary dynamic behavioural features.

\textbf{Refined MI estimation for binary features.} Re-run the Section~\ref{sec:mi} MI analysis with a mixed discrete--continuous estimator (Ross 2014 or Gao et al.~2017) for the 9 binary features in PRISM. Low-cost refinement that improves the precision of the separability table without changing qualitative conclusions.

\textbf{Adversarial robustness evaluation.} As PRISM-based classifiers are developed, evaluating their robustness to adversarial PE manipulation becomes important. The structural information PRISM preserves may make positional-aware models more robust to manipulations that 1D models miss, or conversely may expose new attack surfaces.

\textbf{Cross-corpus and concept-drift evaluation.} Train on the PRISM corpus and evaluate on contemporary samples from MalwareBazaar 2025--2026, or evaluate temporal drift within the corpus using the BODMAS timestamps. Drift evaluation rewards representations that preserve generalisable structure.

\textbf{BODMAS-equivalent re-collection.} Acquire raw PE binaries for samples temporally equivalent to BODMAS (2019--2020), enabling sample-matched cross-representation comparison on BODMAS-era malware. The current EMBER-compatible sub-corpus excludes BODMAS because of the binary-availability constraint; closing that gap would extend the cross-representation comparison to the most temporally-controlled benchmark available.

\textbf{Provenance-controlled evaluation.} Construct a source-balanced split in which benign and malware samples are drawn from a shared feed, and train an auxiliary probe to predict \emph{source} rather than \emph{label}, in order to bound the residual provenance and disarming signal identified in Section~\ref{sec:results}. This is a prerequisite for any absolute detection claim from the corpus and is independent of the (provenance-robust) controlled comparisons reported here.

\textbf{Multi-seed confirmation on the expanded corpus.} The 20-seed paired analysis (with McNemar tests) is reported here for the non-BODMAS sub-corpus; extending it to the full BODMAS-inclusive 48{,}825-sample corpus, and adding a two one-sided test (TOST) against a pre-registered tail margin, would complete the statistical characterisation of the small EMBER advantage observed across splits.

\section*{Acknowledgements}

The authors thank the teams behind EMBER, BODMAS, SOREL-20M, MalwareBazaar (abuse.ch), and VirusShare for making their datasets available to the research community. BODMAS feature vectors were provided by Limin Yang and Gang Wang (University of Illinois Urbana-Champaign) following their data sharing agreement. MalwareBazaar API access was provided by abuse.ch. SOREL-20M benign samples were made available through Sophos and ReversingLabs under the original data sharing terms. The work of A.I.G.-T. is supported by project PID2023-150310OB-I00, funded by MCIU/AEI/FEDER UE, Spain. 

\bibliographystyle{IEEEtran}
\bibliography{references}

\end{document}